\documentclass[aps,prd,longbibliography,nofootinbib]{revtex4-2}
\makeatletter\def\switch@array{}\makeatother
\usepackage{graphicx}
\usepackage{amsmath,amssymb,bm}
\usepackage{siunitx}
\DeclareSIUnit\year{yr}
\usepackage{booktabs}
\usepackage{microtype}
\usepackage{natbib}
\usepackage{hyperref}
\hypersetup{colorlinks=true,allcolors=blue}
\usepackage{array,tabularx}
\usepackage{makecell}
\usepackage{threeparttable}
\usepackage{multirow}
\usepackage{mathtools}
\newcolumntype{Y}{>{\raggedright\arraybackslash}X}
\newcolumntype{L}{>{\raggedright\arraybackslash}X}
\usepackage{comment}

\newcommand{\yr}{\mathrm{yr}}

\newcommand{\kms}{\mathrm{km\,s^{-1}}}
\newcommand{\Rearth}{R_{\oplus}}

\newcommand{\mars}{\mathrm{Mars}}
\newcommand{\ve}{v_{\mathrm{esc},\oplus}}
\newcommand{\vm}{v_{\mathrm{esc},\mars}}

\newcommand{\uinf}{u_{\infty}}
\newcommand{\vinfearth}{v_{\infty,\oplus}}
\newcommand{\qspec}{q_{\mathrm{k}}}
\newcommand{\fg}{f_{\mathrm{g}}}
\newcommand{\Phard}{P_{\mathrm{hard}}}
\newcommand{\Psoft}{P_{\mathrm{soft}}}

\newcommand{\Pejbio}{P_{\mathrm{ej}}^{\mathrm{bio}}}
\newcommand{\Psys}{P_{\mathrm{esc,sys}}}
\newcommand{\Ptr}{P_{\mathrm{tr}}}
\newcommand{\Pearth}{P_{\oplus}}
\newcommand{\Pentry}{P_{\mathrm{entry}}}
\newcommand{\Pimpact}{P_{\mathrm{impact}}}
\newcommand{\Pest}{P_{\mathrm{est}}}
\newcommand{\Ktr}{K_{\mathrm{tr}}}
\newcommand{\Kchem}{K_{\mathrm{chem}}}
\newcommand{\etal}{\textit{et al.}}
\newcommand{\alphath}{\alpha_{\mathrm{th}}}
\newcommand{\dth}{\delta_{\mathrm{th}}}
\newcommand{\pc}{p_{\mathrm{c}}}

\newcommand{\dent}{d_{\mathrm{ent}}}
\newcommand{\drad}{d_{\mathrm{rad}}}
\newcommand{\dmin}{d_{\min}}
\newcommand{\Fbur}{\mathcal{F}_{\mathrm{bur}}}
\newcommand{\Rmaxsh}{R_{\max,\mathrm{sh}}}
\newcommand{\mumat}{\bm{\mu}}

\newcommand{\XAN}{\mathcal{C}}
\newcommand{\Bbench}{\widetilde{\mathcal{B}}}

\usepackage{pgfplots}
\usepgfplotslibrary{fillbetween}
\pgfplotsset{compat=1.18}

\begin{document}

\title{Natural Lithopanspermia to Earth: Transport, Shielding, and \\ Survival Limits for Solar-System and Extrasolar Donor Classes}

\author{Slava G. Turyshev}
\affiliation{Jet Propulsion Laboratory, California Institute of Technology,\\
4800 Oak Grove Drive, Pasadena, CA 91109-0899, USA}
\date{\today}

\begin{abstract}
Natural lithopanspermia to Earth is a transport-and-establishment hypothesis, not a theory of abiogenesis. We formulate the Earth-specific problem as a semi-analytic screening calculation: after launch, planetary-system escape, transit, Earth interception, atmospheric entry, terminal loading, and post-delivery establishment are imposed, which nonterrestrial donor classes retain a non-negligible transport-survival weight relative to indigenous terrestrial origin? The calculation is not a closed posterior for terrestrial origin, because the abiogenesis prior, donor-life prior, inhabited-ejecta yield, and establishment probability are not known. It instead gives a transport-ranked viability hierarchy and identifies the physical regimes in which transfer or survival is plausible. The framework combines a donor-dependent Earth-directed kernel \(\mathcal{K}_{\oplus}^{(i)}\), a minimum protected-depth envelope \(d_{\min}(t_{\mathrm{fl}})\), and a survival-weighted buried-volume fraction \(\Fbur\) for low-shock spall fragments. The relevant carriers are treated as lithic impact ejecta whose density, porosity, volatile/mineral content, thermal diffusivity, shielding column, fragmentation behavior, and shock history enter the survival operators rather than being ignored. This separation makes clear what is derived analytically, what is adopted from experiments or dynamical calculations, and what remains a target for hydrocode, radiation-transport, entry, and biological-survival modeling. For Earth's origin history, hard panspermia is transport-competitive only on Solar-System scales. Early Mars combines demonstrated lithic exchange with Earth, low escape speed, early aqueous habitability, and a rare \(10^{2}\)--\(10^{4}\,\mathrm{yr}\) transfer tail; the more common \(10^{5}\)--\(10^{7}\,\mathrm{yr}\) martian transfer regime requires lightly shocked meter-class carriers and is strongly penalized by \(\Fbur\). Beyond the Solar System, low-\(\uinf\) Solar-System capture, long-duration radiation survival, and early donor timing fail jointly; even the Solar birth-cluster channel gives only a strongly suppressed benchmark composite seeding coefficient. Indigenous terrestrial origin therefore remains the transport null, early Mars is the only quantitatively serious external hard-panspermia alternative, and extrasolar or intergalactic hard panspermia is not competitive for Earth's actual origin history under current constraints. Soft panspermia is a separate and much stronger claim: exogenous delivery of organics, minerals, and catalytic feedstock plausibly enriched early terrestrial prebiotic chemistry without importing living cells.

\end{abstract}

\maketitle

\section{Introduction}
\label{sec:intro}

The origin of terrestrial life can no longer be posed simply as a choice between ``abiogenesis on Earth'' and ``life elsewhere.'' Several empirical developments now make the question quantitative. Earth hosted life very early, so any exogenous seeding channel must compete with a fast terrestrial biosphere clock \cite{dodd2017}. Rocky exchange within the inner Solar System is real: martian meteorites show that launch from Mars, interplanetary transfer, atmospheric entry, and deposition on Earth all occur in nature \cite{gladman1996,gladman1997,herd2024}. Earth also was not chemically closed: extraterrestrial delivery of organics, minerals, and catalytic feedstock is directly established by meteorites and returned samples \cite{oba2022,glavin2025,furukawa2026}. Finally, modern calculations of launch shock, space exposure, radiation transport, and low-\(\uinf\) Solar-System capture now constrain which donor classes are physically credible and which are not \cite{adams2022,dehnen2022,cao2024}.

The purpose of this paper is to construct a semi-analytic Earth-directed screening model for natural lithopanspermia. The model does not attempt to decide whether life began on Earth or elsewhere by assigning a closed posterior probability, nor does it simulate a single specified impact event. Instead, it asks which donor classes survive the same sequence of physical filters: biological launch, donor-system escape, transit survival, Earth interception, atmospheric entry, terminal loading, and post-delivery establishment. The quantities derived below are the conditional transport kernel, the protected-depth requirement, the survival-weighted fraction of low-shock carrier volume, the martian fast/slow transfer penalty, and order-of-magnitude stress and timescale bounds. The quantities adopted from the literature are the experimental shock-survival benchmarks, martian transfer-time statistics, radiation-shielding depths, entry-heating constraints, and extrasolar capture/seeding coefficients. The quantities not modeled from first principles are detailed impact hydrodynamics, lithology-specific chemistry, organism-specific survival curves, atmospheric fragmentation histories, target-dependent impact mechanics, and post-arrival ecology. This separation is central: the paper uses analytical filters to determine where panspermia remains physically plausible, not to replace detailed numerical or experimental work.

Panspermia is often discussed as though it were a single proposition. It is not. At minimum one must distinguish \emph{hard panspermia}, in which viable replicators are transported and later establish a biosphere, from \emph{soft panspermia}, in which the delivered material consists of organics, minerals, or catalytic feedstock that may assist abiogenesis without importing living cells. That distinction is not semantic; the two channels have very different physical requirements and very different evidentiary status.

The core problem is conjunctive. For hard panspermia, we write
\begin{equation}
\Phard = P_{\mathrm{life}}\,\Ktr\,\Pest,
\label{eq:phard}
\end{equation}
where \(P_{\mathrm{life}}\) is the probability that the donor world actually hosts life, \(\Ktr\) is the purely transport kernel, and \(\Pest\) is the conditional probability that material arriving on Earth actually establishes a terrestrial biosphere. The transport kernel is
\begin{equation}
\Ktr = \Pejbio\,\Psys\,\Ptr\,\Pearth\,\Pentry\,\Pimpact,
\label{eq:ktr}
\end{equation}
where \(\Pejbio\) is the probability that impact ejecta contain biologically viable material and leave the planetary surface without sterilization, \(\Psys\) is the probability of escape from the donor planetary system, \(\Ptr\) is the probability of surviving transit, \(\Pearth\) is the probability of Earth interception, \(\Pentry\) is the probability of surviving atmospheric entry, and \(\Pimpact\) is the probability of surviving the terminal deceleration and impact environment.

Eq.~(\ref{eq:ktr}) should be read as a compact chain-rule notation, not as an assumption that the stages are statistically independent. More explicitly, if \(E_{\mathrm{ej}}\), \(E_{\mathrm{esc}}\), \(E_{\mathrm{tr}}\), \(E_{\oplus}\), \(E_{\mathrm{ent}}\), and \(E_{\mathrm{imp}}\) denote successful biological launch, system escape, transit, Earth interception, atmospheric entry, and terminal-impact survival, then
\begin{align}
\Ktr
&=
P(E_{\mathrm{ej}})
P(E_{\mathrm{esc}}\mid E_{\mathrm{ej}})
P(E_{\mathrm{tr}}\mid E_{\mathrm{ej}},E_{\mathrm{esc}})
\nonumber\\
&\qquad\times
P(E_{\oplus}\mid E_{\mathrm{ej}},E_{\mathrm{esc}},E_{\mathrm{tr}})
P(E_{\mathrm{ent}}\mid E_{\mathrm{ej}},E_{\mathrm{esc}},E_{\mathrm{tr}},E_{\oplus})
\nonumber\\
&\qquad\times
P(E_{\mathrm{imp}}\mid E_{\mathrm{ej}},E_{\mathrm{esc}},E_{\mathrm{tr}},E_{\oplus},E_{\mathrm{ent}}).
\label{eq:ktr_conditional}
\end{align}
The notation in Eq.~(\ref{eq:ktr}) is therefore a stage bookkeeping device. Correlations among carrier radius, burial depth, shock pressure, flight time, and encounter speed are retained in the state-space kernel below.

For soft panspermia, a different factorization is more appropriate:
\begin{equation}
\Psoft = P_{\mathrm{org}}\,\Kchem\,P_{\mathrm{assist}},
\label{eq:psoft}
\end{equation}
where \(P_{\mathrm{org}}\) is the probability that the donor population contains chemically relevant feedstock, \(\Kchem\) is the transport kernel for that feedstock, and \(P_{\mathrm{assist}}\) is the probability that the delivered material nontrivially assists terrestrial prebiotic chemistry. The hard and soft channels should therefore never be conflated in a technical discussion.

The analysis below should therefore be read as a transport-survival screen. Its output is not an absolute probability for terrestrial origin, but a ranked set of donor-class weights conditioned on the existence of suitable source material. The quantitative contribution is fourfold. First, the paper writes the problem as a donor-dependent Earth-directed kernel \(\mathcal{K}_{\oplus}^{(i)}\) that keeps launch, transfer, radiation, interception, entry, and impact survival as separate factors. Second, it introduces the protected-depth envelope \(d_{\min}(t_{\mathrm{fl}})\), which maps flight time into a minimum carrier architecture. Third, it introduces the survival-weighted buried-volume fraction \(\Fbur\), which asks what fraction of the biologically loaded low-shock ejecta population remains protected after the joint entry-and-radiation requirement is imposed. Fourth, it isolates the martian architecture-and-timing penalty \(\eta_{\mars}\), which separates the biologically privileged fast Mars-to-Earth tail from the more common Myr-scale martian-meteorite channel.

Two empirical facts make the problem Earth-specific. First, rocky exchange within the inner Solar System is real: martian meteorites demonstrate that launch from Mars, interplanetary transfer, atmospheric entry, and deposition on Earth occur in nature. Second, Earth hosted life very early. Any successful exogenous-seeding scenario must therefore operate early enough, and with sufficient survival probability, to compete with Earth's own rapid acquisition of biology. The claim advanced below is consequently limited but testable: once the full Earth-directed filter is imposed, the hard-panspermia competition collapses to terrestrial origin versus early Mars, while extrasolar and intergalactic hard panspermia remain physically possible but transport-noncompetitive. Soft panspermia is treated separately as chemical enrichment rather than imported biology.

The paper is organized as follows. Section~\ref{sec:framework} defines the Earth-directed transport kernel and the adopted viability envelopes. Section~\ref{sec:launch} treats launch from donor worlds and donor prerequisites. Sections~\ref{sec:solarsystem} and \ref{sec:extrasolar} analyze transfer within and beyond the Solar System. Section~\ref{sec:biology} develops the transit-survival and carrier-architecture constraints. Section~\ref{sec:arrival} treats Earth interception, atmospheric entry, and terminal loading. Section~\ref{sec:donor} synthesizes the integrated donor hierarchy. Section~\ref{sec:establishment} treats post-delivery establishment and soft panspermia. Section~\ref{sec:discussion} isolates the decisive unknowns and falsifiers, and Sec.~\ref{sec:concl} summarizes the conclusions.

\section{Earth-directed transport kernel and adopted viability envelopes}
\label{sec:framework}

The central mistake in casual panspermia arguments is to reason from the total kinetic energy of an impactor rather than from the stage-specific pressure, heating, and dose histories of the actual biological microenvironment inside the carrier rock. A carrier can survive one stage of the chain and fail at another, and each stage imposes a different geometric requirement. We therefore formulate the problem in terms of a donor-class dependent Earth-directed transport-survival kernel.

For donor class \(i\), let the launch-state vector be
\[
\xi\equiv(R,d,p,\mumat),
\]
where \(R\) is carrier radius, \(d\) is the burial depth of the biological cargo, \(p\) is the peak launch pressure experienced by the relevant microenvironment, and \(\mumat\) denotes the carrier material state. In this paper \(\mumat\) is not a new free probability factor; it is a compact way to keep track of lithology-dependent properties such as bulk density, porosity, grain size, thermal diffusivity, volatile and organic content, hydration state, redox chemistry, and mechanical cohesion. Let the encounter-state vector be donor-dependent. For Solar-System donors,
\[
\zeta_i\equiv(t_{\mathrm{fl}},\vinfearth),
\]
where \(t_{\mathrm{fl}}\) is donor-to-Earth flight time and \(\vinfearth\) is the geocentric asymptotic speed of a carrier already on an Earth-encounter trajectory. For extrasolar donors, the upstream Solar-System capture problem introduces an additional heliocentric incoming speed \(\uinf\), so that one may write
\[
\zeta_i\equiv(t_{\mathrm{fl}},\uinf,\vinfearth)
\]
for the captured interstellar channel. The two speeds should not be identified: \(\uinf\) controls whether an interstellar object can be captured by the Solar System, whereas \(\vinfearth\) controls Earth interception and entry speed after the object has entered the relevant Earth-encounter phase space. We define
\begin{equation}
\Gamma_i(\xi)\equiv
\frac{dN_{\mathrm{ej,bio}}^{(i)}}{d\xi},
\qquad
N_{\mathrm{ej,bio}}^{(i)}=
\int d\xi\,\Gamma_i(\xi),
\label{eq:Gamma_norm}
\end{equation}
so that \(\Gamma_i\) is a differential count of biologically loaded ejecta in launch-state space, not a dimensionless probability. Throughout this paper, \(N_{\mathrm{ej,bio}}^{(i)}\) is interpreted as the expected biologically loaded ejecta yield \emph{conditional on donor class \(i\) actually hosting life}. The separate factor \(P_{\mathrm{life}}^{(i)}\) below therefore represents the prior probability that the donor is inhabited at all, while \(N_{\mathrm{ej,bio}}^{(i)}\) counts inhabited ejecta given that condition. This convention prevents double-counting donor-life uncertainty. Equivalently,
\[
\phi_i(\xi)\equiv \Gamma_i(\xi)/N_{\mathrm{ej,bio}}^{(i)}
\]
is the normalized launch-state distribution conditional on an inhabited donor.

The normalized Earth-directed transport-survival kernel is then
\begin{align}
\mathcal{K}_{\oplus}^{(i)}
&=
\int d\xi\,\phi_i(\xi)\,
S_{\mathrm{sh}}(p;\mumat)\,
P_{\mathrm{esc}}^{(i)}(\xi)
\int d\zeta_i\,
P_{\mathrm{dyn}}^{(i)}(\zeta_i\mid\xi,\mathrm{esc})
\nonumber\\
&\qquad\qquad\times
S_{\mathrm{rad}}(R,d,t_{\mathrm{fl}};\mumat)\,
P_{\oplus}(\vinfearth)\,
S_{\mathrm{ent}}(R,d,v_{\mathrm{imp}};\mumat)\,
S_{\mathrm{imp}}(R,d,v_{\mathrm{imp}};\mumat).
\label{eq:kearth}
\end{align}
Here \(P_{\mathrm{dyn}}^{(i)}(\zeta_i\mid\xi,\mathrm{esc})\) is a subnormalized conditional density over the appropriate encounter variables for donor class \(i\). Its integral over \(\zeta_i\) is the probability that an escaped carrier with launch state \(\xi\) reaches the relevant Earth-encounter phase space. For extrasolar donors, this density includes the upstream low-\(\uinf\) Solar-System capture filter; for all donors, \(P_{\oplus}(\vinfearth)\) is the conditional Earth-interception probability for the resulting geocentric encounter class, including gravitational focusing. The survival operators \(S_{\mathrm{sh}},S_{\mathrm{rad}},S_{\mathrm{ent}},S_{\mathrm{imp}}\in[0,1]\) encode launch-shock survival, radiation survival, atmospheric-entry survival, and terminal-impact survival, respectively, and are allowed to depend on carrier composition and structure through \(\mumat\). Written this way, \(\mathcal{K}_{\oplus}^{(i)}\) is dimensionless and explicitly conditional: correlations among \(R\), \(d\), \(p\), \(\mumat\), \(t_{\mathrm{fl}}\), \(\uinf\), and \(\vinfearth\) enter through \(\phi_i(\xi)\) and \(P_{\mathrm{dyn}}^{(i)}(\zeta_i\mid\xi,\mathrm{esc})\), rather than being assumed away.

The kernel \(\mathcal{K}_{\oplus}^{(i)}\) is therefore not itself an event count. Conditional on donor class \(i\) hosting life, the expected number of successful seed deliveries represented by this channel is
\[
N_{\mathrm{seed}\mid\mathrm{life}}^{(i)}
=
N_{\mathrm{ej,bio}}^{(i)}
\mathcal{K}_{\oplus}^{(i)}
P_{\mathrm{est}}^{(i\rightarrow\oplus)} .
\]
Multiplication by \(P_{\mathrm{life}}^{(i)}\) gives the corresponding unnormalized channel weight used in Sec.~VIII. This convention keeps the transport kernel dimensionless and prevents source-biology terms from being absorbed into \(\mathcal{K}_{\oplus}^{(i)}\).

Eq.~(\ref{eq:kearth}) is a screening kernel, not a fully evaluated end-to-end Monte Carlo model. A closed evaluation would require coupled impact-spall hydrodynamics for \(\phi_i(R,d,p,\mumat)\), orbital integrations for \(P_{\mathrm{dyn}}^{(i)}(t_{\mathrm{fl}},\uinf,\vinfearth\mid \xi,\mathrm{esc})\), radiation transport for \(S_{\mathrm{rad}}\), atmospheric-entry and fragmentation calculations for \(S_{\mathrm{ent}}\), and target-dependent impact modeling for \(S_{\mathrm{imp}}\), all coupled to organism- and lithology-specific survival curves. The present work instead uses conservative, literature-anchored viability envelopes to identify which donor classes are excluded or remain competitive at the order-of-magnitude level. Numerical values below should therefore be interpreted as screening thresholds and benchmark penalties, not as precise posterior probabilities.

The boundary between calculation and assumption is explicit. Derived analytically here are the Earth-focusing and entry-speed scalings, the protected-depth condition, the conductive thermal-depth estimate, the terminal stopping-length scaling, the buried-volume fraction \(\Fbur\), and the martian architecture penalty \(\eta_{\mars}\). Adopted from prior experiments and dynamical studies are the benchmark launch-shock range, entry-heating floor, radiation-shielding envelope, Mars-transfer windows, and extrasolar capture/seeding coefficients. Not modeled here are mineral-specific hydrocodes, microbial repair kinetics, pore-scale chemistry, coupled entry breakup, or target-specific impact survival. Those omissions do not make the calculation unconstrained; they define the parameters whose values must be supplied by future simulations or experiments to move a donor class across the viability boundaries.

For hard panspermia, three stage-specific screening scales dominate the present donor ranking. Viable low-shock launch is concentrated below
\begin{equation}
p \lesssim p_{\mathrm{sh}},
\qquad
p_{\mathrm{sh}}\sim 1\text{--}3~\mathrm{GPa},
\label{eq:psh}
\end{equation}
for a direct modern \textit{D.~radiodurans} benchmark, with older work showing a broader upper tail into the multi-GPa to few-tens-of-GPa regime for selected spores and lichens \cite{stoffler2007,horneck2008,zhao2026}. Earth-entry survival requires burial depths \cite{fajardo2005,delatorre2010,horneck2012,cottin2017}
\begin{equation}
d \gtrsim d_{\mathrm{ent}},
\qquad
d_{\mathrm{ent}}\sim 0.02\text{--}0.05~\mathrm{m},
\label{eq:entryfloor}
\end{equation}
and efficient capture of an interstellar object by the Solar System \cite{dehnen2022} is concentrated at low heliocentric incoming speed,
\begin{equation}
\uinf \lesssim v_{\mathrm{cap}},
\qquad
v_{\mathrm{cap}}\sim 4~\kms.
\label{eq:vcap}
\end{equation}
These scales are not universal biological constants. They are benchmark filters: donor classes that generically violate one or more of them are transport-disfavored before any donor-life prior is assigned.

The radiation filter is explicitly time dependent, but no single function \(d_{\mathrm{rad}}(t_{\mathrm{fl}})\) is a universal survival law. The required protected depth depends on organism, metabolic state, hydration, repair after rehydration, rock composition, shielding geometry, galactic cosmic-ray environment, solar energetic-particle history, and the minimum surviving population required for establishment. We therefore use the following as a conservative screening envelope against false-positive long-transfer claims, not as a sharp sterilization boundary:
\begin{equation}
d_{\mathrm{rad}}(t_{\mathrm{fl}}) \sim
\begin{cases}
10^{-3}\text{--}10^{-2}~\mathrm{m}, & t_{\mathrm{fl}} \lesssim 10~\yr, \\
10^{-2}\text{--}10^{-1}~\mathrm{m}, & 10^{2}\,\yr \lesssim t_{\mathrm{fl}} \lesssim 10^{4}\,\yr, \\
0.5\text{--}1~\mathrm{m}, & 10^{5}\,\yr \lesssim t_{\mathrm{fl}} \lesssim 10^{7}\,\yr, \\
\gg 1~\mathrm{m}, & t_{\mathrm{fl}} \gtrsim 10^{8}\,\yr.
\end{cases}
\label{eq:drad}
\end{equation}
The operative protected-depth requirement is
\begin{equation}
d_{\min}(t_{\mathrm{fl}})=\max\!\left[d_{\mathrm{ent}},\,d_{\mathrm{rad}}(t_{\mathrm{fl}})\right].
\label{eq:dmin}
\end{equation}
Eqs.~(\ref{eq:drad}) and (\ref{eq:dmin}) should therefore be interpreted as a scenario-ranking device. They are conservative with respect to excluding long, poorly shielded hard-panspermia channels, but they are not conservative in every biological sense: for example, a surface-biased biosphere, a small founder population, or limited post-arrival repair would make the survival requirement stricter. The robust conclusion is the scale break: the rare \(10^{2}\)--\(10^{4}\,\yr\) Mars-to-Earth tail remains compatible with centimeter-to-decimeter protected depths, whereas the common \(10^{5}\)--\(10^{7}\,\yr\) martian-meteorite regime naturally requires protected depths of order \(0.5\)--\(1~\mathrm{m}\).

\begin{figure}[t]
\centering
\includegraphics[width=0.60\columnwidth]
{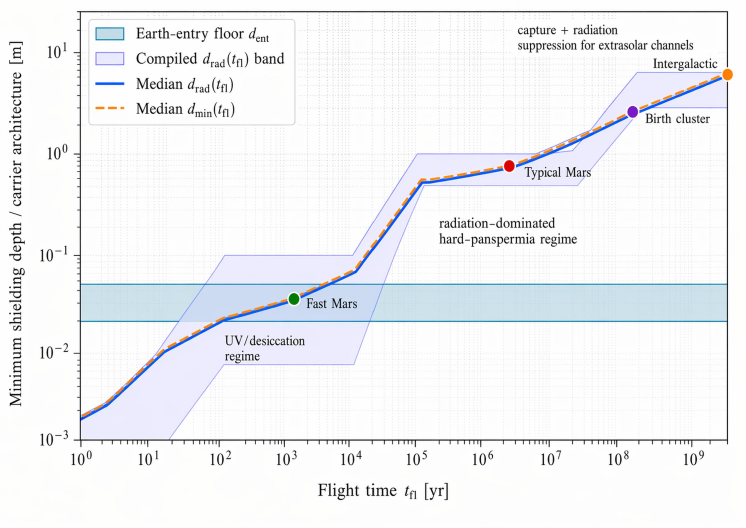}
\caption{Minimum protected depth \(d_{\min}(t_{\mathrm{fl}})=\max[d_{\mathrm{ent}},d_{\mathrm{rad}}(t_{\mathrm{fl}})]\) as a function of donor-to-Earth flight time \(t_{\mathrm{fl}}\). The horizontal floor is set by Earth entry, while the rising branch is set by cumulative radiation dose. The band is a literature-anchored screening envelope, not a universal survival curve; organismal repair, mineralogy, porosity, shielding geometry, and required founder population would move the boundaries. Scenario markers show the scale separation directly: the rare fast Mars tail remains compatible with centimeter-to-decimeter carriers, whereas the common \(10^{5}\)--\(10^{7}\,\yr\) martian channel already requires meter-class shielding and extrasolar channels are penalized even before capture is considered.}
\label{fig:dmin}
\end{figure}

Figure~\ref{fig:dmin} makes the scale break explicit: fast transfer is compatible with centimeter-to-decimeter carriers, whereas Myr-scale hard panspermia naturally drives the problem into the meter-class-boulder regime.

At launch, the relevant bulk energy scale is the specific kinetic energy,
\begin{equation}
\qspec = \frac{v^2}{2},
\label{eq:qspec}
\end{equation}
where \(v\) is the ejection speed. For Mars, \(\vm \approx \SI{5.0}{\kms}\), so \(\qspec \approx \SI{12.5}{MJ\,kg^{-1}}\); for Earth, \(\ve \approx \SI{11.2}{\kms}\), so \(\qspec \approx \SI{62.7}{MJ\,kg^{-1}}\). Those values look sterilizing if interpreted as uniform thermal loading, but the relevant biological state variable is the nonuniform pressure-temperature history of the actual carrier fragment, not the bulk kinetic energy.

At Earth interception, gravitational focusing modifies the collision cross section to
\begin{equation}
\sigma_{\oplus} = \pi \Rearth^2 \Big( 1 + \frac{\ve^2}{\vinfearth^2} \Big) = \pi \Rearth^2 \fg,
\label{eq:focusing}
\end{equation}
where \(\fg\) is the focusing factor. The resulting entry speed at the top of the atmosphere is
\begin{equation}
v_{\mathrm{imp}} = \sqrt{\vinfearth^2 + \ve^2}.
\label{eq:vimp}
\end{equation}
Low \(\vinfearth\) both increases Earth-interception probability and reduces the incremental dynamical load beyond Earth's unavoidable gravitational acceleration.

During transit, the cumulative dose can be written schematically as
\begin{equation}
D_{\mathrm{tot}}(t) = \int_0^t \dot D_{\mathrm{GCR}}(R,d)\,dt' + \sum_j D_j^{\mathrm{transient}},
\label{eq:dose}
\end{equation}
where \(\dot D_{\mathrm{GCR}}\) is the internal dose rate from background galactic cosmic rays and the discrete terms represent rare transients such as nearby supernovae or gamma-ray bursts. Survival requires \(D_{\mathrm{tot}}<D_{\mathrm{ster}}\), with \(D_{\mathrm{ster}}\) depending on the organism, hydration state, and the possibility of post-arrival repair. The most important practical point is that entry heating and long-duration radiation do not impose the same architectural requirement: entry sets a centimeter-scale burial floor, whereas Myr survival drives the carrier toward meter-scale shielding.

The biological stress factors can be written as a conditional gate rather than as a universal survival constant:
\begin{equation}
S_{\mathrm{bio}}
= P\!\left[\mathcal{G}_{\mathrm{sh}}\cap\mathcal{G}_{\mathrm{heat}}\cap
\mathcal{G}_{\mathrm{rad}}\cap\mathcal{G}_{\mathrm{des}}\cap
\mathcal{G}_{\mathrm{chem}}\cap\mathcal{G}_{\mathrm{frag}}\mid \xi,\zeta_i\right].
\label{eq:stress_gate}
\end{equation}
Here the gates represent, respectively, tolerable shock pressure and shock heating, peak temperature and heating duration, cumulative ionizing dose, vacuum/UV/desiccation exposure, chemical alteration of the microenvironment, and fragmentation or mechanical disruption. Eq.~(\ref{eq:stress_gate}) is not an independence assumption. It is a compact statement of what the screening operators in Eq.~(\ref{eq:kearth}) must physically represent.

\begin{table*}[t]
\caption{Quantitative constraints on natural panspermia to Earth. Values summarize the most relevant contemporary dynamical, experimental, and flight results for hard panspermia, together with the conservative screening use made of them in this work.}
\label{tab:keyfilters}
\centering
\setlength{\tabcolsep}{2pt}
\renewcommand{\arraystretch}{1.08}
\begin{tabular}{@{}p{0.13\textwidth} p{0.17\textwidth} p{0.33\textwidth} p{0.35\textwidth}@{}}
\toprule
Stage & Representative result & Quantitative constraint from the literature & Conservative use in the present paper \\
\midrule
Carrier material and chemistry & Meteorite physical-property surveys, martian meteorites, rover geochemistry, and returned-sample organics \cite{britt2003,macke2011,udry2020,hurowitz2025,oba2022,glavin2025} & Relevant carriers range from coherent igneous or sedimentary silicates and regolith breccias to more porous, volatile-rich carbonaceous material; porosity, hydration, organics, and mineral redox state affect shielding, microhabitats, devolatilization, and fragmentation & These properties are collected in the material vector \(\mumat\); the fiducial hard-panspermia carrier is a lithified, endolithic silicate or breccia fragment, while volatile-rich carbonaceous material is treated mainly as a soft-panspermia source unless it is sufficiently coherent and shielded \\
Impact ejection & Spallation theory and hypervelocity experiments \cite{melosh1984,mastrapa2001,stoffler2007,horneck2008,zhao2026} & Low-shock ejecta can be launched from Mars; direct modern tests show \textit{D.~radiodurans} survival up to about \SI{3}{GPa}, while selected older spore/lichen experiments support a broader viable tail into the multi-GPa to few-tens-of-GPa regime & Viable launch is treated as a low-shock tail, not a property of the bulk ejecta population; conservative screening uses \(p_{\mathrm{sh}}\sim 1\text{--}3\) GPa as a direct benchmark and recognizes a broader experimental upper tail \\
Transit time & Martian-meteorite exposure ages and transfer calculations \cite{gladman1997,mileikowsky2000} & Measured martian meteorites record about \(0.35\) to \(16\) Myr exposure ages; roughly \(1\%\) of ejecta can reach Earth within \SI{16000}{yr} and \(10^{-4}\) within \SI{100}{yr} & Fast Mars transfer is treated as the biologically privileged channel; long-lived meteorites are dynamically common but biologically far less favorable \\
Vacuum and UV & Tanpopo ISS exposure \cite{kawaguchi2020} & Dried \textit{D.~radiodurans} pellets \(\ge 500\,\mu\mathrm{m}\) survived 3 years in space; \(1\) mm pellets correspond to estimated survival times of about \(2\) to \(8\) years in interplanetary space & UV/desiccation are treated as manageable only for shielded aggregates or lithic carriers on short flights; naked-cell interstellar transfer is not treated as viable \\
Ionizing radiation & Radiation-transport models in rock \cite{mileikowsky2000,dartnell2011} & Rocks smaller than about \(60\) cm can have enhanced central dose from secondaries; effective external shielding begins for boulders of diameter \(\gtrsim 2.6\) m; at burial depths of order \(1\) m, sterilizing doses accumulate on Myr timescales & Radiation is taken as the dominant long-flight bottleneck; centimeter-to-decimeter shielding is assigned only to fast Mars transfer, whereas Myr survival requires meter-class carriers \\
Atmospheric entry & Re-entry and recovery experiments \cite{fajardo2005,delatorre2010,horneck2012,cottin2017} & Shallow burial can be sterilized down to roughly \(2\) to \(5\) cm, while deeper interiors can remain cool enough to survive favorable entries & Entry survival is screened by a burial floor \(d_{\mathrm{ent}}\sim 0.02\text{--}0.05\) m rather than by total impact energy \\
Terminal impact and fragmentation & Atmospheric disruption, entry, and impact-survival constraints \cite{bland2003,fajardo2005,delatorre2010,dartnell2011} & Earth arrival occurs at \(v_{\mathrm{imp}}\ge v_{\mathrm{esc},\oplus}\); contact-zone stresses can be enormous, but transmitted interior stress depends on fragmentation, sacrificial layers, obliquity, and target compliance & Terminal survival is treated as a severe conditional filter controlled by local pressure-temperature history and stopping length, not by bulk kinetic energy alone \\
Interstellar capture & Solar-System capture calculations \cite{adams2022,dehnen2022,cao2024} & Earth-seeding expectations are about \(3\times10^{-5}f_{\mathrm{seed}}\) in the birth cluster and about \(5\times10^{-10}f_{\mathrm{seed}}\) in the Galactic field; Solar-System capture is dominated by \(\uinf < 4\,\kms\) & Extrasolar transfer is screened by the combined requirements of \(\uinf\lesssim 4\,\kms\), long-term radiation survival, and early donor timing; only the birth-cluster channel survives as a nonzero but strongly suppressed possibility \\
\bottomrule
\end{tabular}
\end{table*}

\begin{table*}[t]
\caption{Dominant free inputs and sensitivity directions in the screening model. The table separates quantities constrained mainly by transport or materials physics from quantities that remain biological or geochemical priors.}
\label{tab:sensitivity_budget}
\centering
\setlength{\tabcolsep}{3pt}
\renewcommand{\arraystretch}{1.08}
\begin{tabular}{@{}p{0.17\textwidth} p{0.18\textwidth} p{0.30\textwidth} p{0.30\textwidth}@{}}
\toprule
Input & Where it enters & Sensitivity direction & Main needed improvement \\
\midrule
\(p_{\mathrm{sh}}\) and launch-pressure distribution &
\(S_{\mathrm{sh}}(p;\mumat)\phi_i(R,d,p,\mumat)\) &
A stricter low-shock limit reduces the viable ejecta tail; a broader low-pressure spall population increases it. &
Impact-spall simulations and shock-survival experiments by lithology. \\

Carrier material state \(\mumat\) &
\(\phi_i\), \(S_{\mathrm{rad}}\), \(S_{\mathrm{ent}}\), and \(S_{\mathrm{imp}}\) &
Higher areal density and lower thermal diffusivity improve shielding; porosity and hydration can create habitats but also increase fragmentation, devolatilization, and chemical-alteration risks. &
Lithology-specific coupling among porosity, volatile content, thermal properties, strength, and biological loading. \\

\(\dmin(t_{\mathrm{fl}})\), \(D_{\mathrm{ster}}\), and repair after rehydration &
\(S_{\mathrm{rad}}\), Eqs.~(\ref{eq:drad})--(\ref{eq:dmin}) &
Larger required shielding moves the viable population to larger \(R\) and suppresses long-transfer channels. &
Radiation transport coupled to survival assays for dormant, shielded communities. \\

\(d_{\mathrm{ent}}\) and entry-fragmentation history &
\(S_{\mathrm{ent}}\), Eq.~(\ref{eq:entryfloor}) &
A larger entry-heating floor suppresses centimeter-scale carriers; fragmentation can either expose interiors or shed sacrificial material. &
Entry simulations tied to recovered-meteorite thermal histories. \\

\(\Rmaxsh\) &
Upper limit of Eq.~(\ref{eq:Fbur}) &
If \(\Rmaxsh\lesssim \dmin\), the low-shock channel closes; if \(\Rmaxsh\) reaches meter-to-multi-meter radii, Myr transfer remains partially open. &
Hydrocode estimates of maximum lightly shocked spall fragments. \\

\(\beta\), \(R_{\min}\), and fragment spectrum shape &
Eq.~(\ref{eq:spectrum}) and Eq.~(\ref{eq:Fbur}) &
Steeper spectra place more biological volume in small carriers and increase the penalty from \(d_{\min}\). &
Size distributions for lightly shocked ejecta, not total ejecta. \\

Biological loading exponent \(\gamma\) or shell depth \(h_{\mathrm{bio}}\) &
Eqs.~(\ref{eq:Fbur_gamma})--(\ref{eq:f_shell}) &
Volume-loaded endolithic cargo is most favorable; shallow-shell loading is strongly penalized and gives zero protected cargo when \(h_{\mathrm{bio}}\le d_{\min}\). &
Constraints on subsurface habitability and depth distribution of biomass. \\

Fast-transfer fraction \(f_{\mathrm{fast}}\) and \(\wp_{\mars}(t_{\mathrm{fl}})\) &
Eqs.~(\ref{eq:wpmars_mix})--(\ref{eq:etamars}) &
A larger fast tail reduces radiation exposure and raises \(\eta_{\mars}\); a slow-dominated distribution requires larger carriers or larger source-side compensation. &
Early-Solar-System Mars-to-Earth transfer ensembles. \\

Low-\(\uinf\) Solar-System capture efficiency &
\(P_{\mathrm{dyn}}^{(i)}\), Eq.~(\ref{eq:vcap}) &
Broadening the efficient-capture window would increase extrasolar weights; under current calculations, the low-\(\uinf\) capture and long-duration radiation penalties act in the same direction. &
Capture simulations for birth-cluster and Galactic-field environments. \\

Geocentric Earth-encounter speed \(\vinfearth\) &
\(P_{\oplus}(\vinfearth)\), \(S_{\mathrm{ent}}\), \(S_{\mathrm{imp}}\), Eqs.~(\ref{eq:focusing})--(\ref{eq:vimp}) &
Lower \(\vinfearth\) increases gravitational focusing and lowers the incremental entry speed, but Earth gravity still imposes \(v_{\mathrm{imp}}\ge v_{\mathrm{esc},\oplus}\). &
Earth-encounter ensembles tied to entry and fragmentation models. \\

\(P_{\mathrm{life}}^{(i)}\), \(N_{\mathrm{ej,bio}}^{(i)}\), and \(P_{\mathrm{est}}^{(i\rightarrow\oplus)}\) &
Definitions of \(\Gamma_i\), \(W_i\), and \(\mathcal{B}_i\) &
These set absolute origin probabilities but are not determined by transport physics. &
Independent constraints on donor biospheres, inhabited ejecta yields, and establishment ecology. \\

\bottomrule
\end{tabular}
\end{table*}

\section{Launch from donor worlds}
\label{sec:launch}

\subsection{Low-shock ejection is the only credible launch channel}

The classical objection to lithopanspermia was simple: material fast enough to escape a planet should be melted, vaporized, or at least strongly sterilized. Melosh's spallation model showed why that objection is incomplete: near-surface fragments can be accelerated by reflected tensile waves while avoiding the extreme peak pressures experienced deeper in the target \cite{melosh1984}. In other words, the biologically relevant fragments are not representative of the average ejecta parcel.

The spallation picture is now supported by experiment. Hypervelocity tests with granite and Mars-like targets show that microorganisms in selected fragments can survive launch-like events \cite{mastrapa2001,stoffler2007,horneck2008}. A recent direct study using \textit{Deinococcus radiodurans} found survival remained high at about \SI{1.4}{GPa}, substantial at about \SI{2.4}{GPa}, and still measurable below about \SI{3}{GPa} \cite{zhao2026}. These values do not imply that all ejecta are biologically viable; they imply only that the viable fraction is concentrated in a low-shock tail. That tail is exactly the one relevant for panspermia.

\subsection{Carrier materials and chemically realistic impactors}

The relevant impactors are not bare cells or generic point masses, but coherent lithic fragments that provide mechanical support, thermal inertia, and radiation shielding. For Mars, the observational analogues are martian meteorites and early-Mars sedimentary or hydrothermal materials: igneous basaltic and ultramafic rocks, breccias, regolith-cemented material, and possible evaporite- or carbonate-bearing substrates associated with aqueous environments \cite{udry2020,herd2024,grotzinger2014,mangold2021,vago2017}. A representative silicate carrier has \(\rho\sim2.5\text{--}3.4\times10^3\,\mathrm{kg\,m^{-3}}\), \(\alpha_{\rm th}\sim10^{-6}\,\mathrm{m^2\,s^{-1}}\), and heat capacity of order \(10^3\,\mathrm{J\,kg^{-1}\,K^{-1}}\). Porosity can improve habitat space and reduce bulk density, but it can also increase fragmentation and shock attenuation in a lithology-dependent way.

Shielding is therefore best thought of as a column-density and geometry problem rather than a generic ``rock size'' problem. Five centimeters of silicate rock corresponds to an areal column \(\rho d\sim1.3\text{--}1.7\times10^2\,\mathrm{kg\,m^{-2}}\), while \(0.5\) m corresponds to \(1.3\text{--}1.7\times10^3\,\mathrm{kg\,m^{-2}}\). Hydrated minerals, salts, clays, carbonates, or ice-bearing pores are biologically attractive because they can concentrate water and solutes, but they are not automatically favorable for transfer: shock devolatilization, oxidative alteration, radiolysis, and entry heating can reduce viability. The present model treats these effects through \(\mumat\), \(\dmin\), and the survival operators rather than by assigning an unconstrained material prior.

\subsection{Donor prerequisites: low escape speed, early habitability, and timing}

A viable donor must satisfy three conditions simultaneously: it must launch lightly shocked material above escape speed, it must host habitable environments early enough for transfer to matter for Earth's origin, and it must do so on trajectories whose subsequent transport kernel is not already negligible. Among known candidates, early Mars best satisfies this conjunction. Its escape speed is only about \SI{5}{\kms}, rocky transfer from Mars to Earth is observationally established, and rover data show that ancient Mars hosted long-lived aqueous environments \cite{grotzinger2014,mangold2021}. By contrast, Venus combines a deeper potential well with a dense atmosphere, and extrasolar donors inherit the additional filters of planetary-system escape, interstellar transfer, and Solar-System capture.

The donor's broader Galactic environment matters mainly through the cumulative radiation background and long-term habitability window. For Mars-to-Earth transfer, those effects are secondary to the much stronger constraints imposed by launch pressure, flight time, shielding depth, and Earth entry. For extrasolar channels they act in the same direction as the dynamical suppression: they do not rescue an already unfavorable transport kernel but reinforce its smallness.

\section{Transfer within the Solar System}
\label{sec:solarsystem}

\subsection{Mars-to-Earth transfer is real; the biological status is the open issue}

The least speculative part of lithopanspermia is the mechanical transfer itself. Martian meteorites on Earth demonstrate that launch from Mars, interplanetary transit, atmospheric entry, and deposition on Earth can all occur in nature \cite{gladman1996,gladman1997,herd2024}. The open question is biological, not dynamical: were any such carriers inhabited, and if so, could any inhabitants survive?

ALH84001 is a useful physical example. Ref.~\cite{weiss2000} inferred that its interior had not exceeded roughly \SI{40}{\celsius} since before ejection from Mars. That specific meteorite is not evidence for life transfer, but it is strong evidence that the interior of at least some martian meteorites remains far colder than naive whole-body energy arguments would suggest.

\subsection{Fast and slow transfer channels}

A crucial distinction separates the most common martian meteorites from the most biologically favorable transfer trajectories. Cosmic-ray exposure ages of known martian meteorites span roughly \(0.35\) to \(16\) Myr \cite{gladman1997}. Those durations are challenging for microbial survival unless the carrier is large enough to provide substantial shielding. Yet trajectory calculations also show that about \(1\%\) of Mars ejecta can reach Earth within \SI{16000}{yr}, and about \(10^{-4}\) can arrive within \SI{100}{yr} \cite{gladman1997,mileikowsky2000}. Those rare short trajectories have a vastly better biological prospect because radiation and desiccation requirements become much less severe.

The resulting synthesis is important: the most biologically relevant carriers need not dominate modern meteorite collections. Surviving, easily recoverable meteorites today are biased toward a particular size and exposure-age distribution. The subset that would have offered the best odds for hard panspermia is likely the rare fast-transfer tail.

\subsection{Earth interception, entry, and terminal loading}
\label{sec:entryimpact}

\begin{figure}[t]
\centering
\includegraphics[width=0.60\columnwidth]{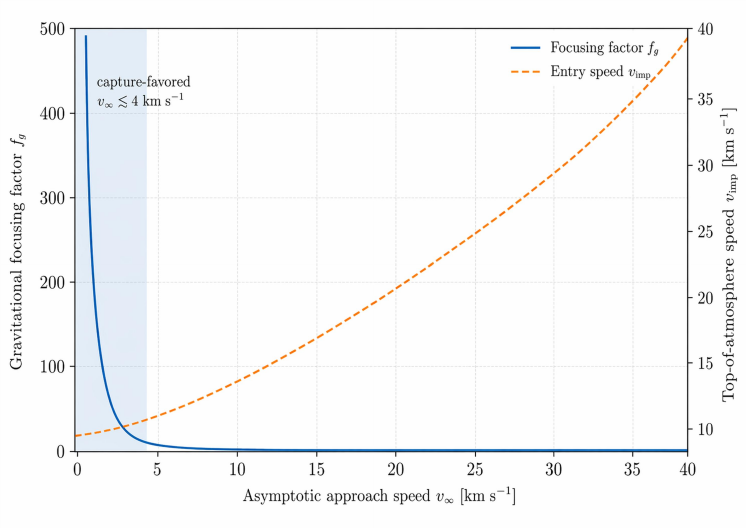}
\caption{Earth-interception kinematics as a function of geocentric asymptotic encounter speed \(\vinfearth\). The solid curve shows the gravitational focusing factor \(\fg=1+\ve^2/\vinfearth^2\), and the dashed curve shows the corresponding top-of-atmosphere speed \(v_{\mathrm{imp}}=\sqrt{\vinfearth^2+\ve^2}\). The shaded low-speed band is a visual reference for the speed scale on which gravitational focusing is large; for extrasolar donors, the separate Solar-System capture filter is controlled by the heliocentric incoming speed \(\uinf\), not by \(\vinfearth\). The physical point is the separation of capture and arrival: low-speed capture and low-speed Earth encounter are related but distinct dynamical conditions, and even the most favorable Earth encounters still reach the atmosphere at meteorite-like speeds because Earth gravity enforces \(v_{\mathrm{imp}}\ge\ve\simeq\SI{11.2}{\kms}\). Survival therefore depends on burial depth, ablation, fragmentation, and target compliance rather than on a mechanically gentle encounter.}
\label{fig:arrivalkin}
\end{figure}

The final orbital step is not purely geometric. Earth gravity enhances the interception cross section by the focusing factor in Eq.~(\ref{eq:focusing}). Figure~\ref{fig:arrivalkin} treats the geocentric Earth-encounter speed \(\vinfearth\), not the upstream heliocentric capture speed \(\uinf\). For Solar-System transfers, \(\vinfearth\) is the natural arrival variable. For extrasolar donors, \(\uinf\lesssim v_{\mathrm{cap}}\) is an upstream Solar-System capture filter, after which a captured object may later acquire some geocentric encounter speed \(\vinfearth\) at Earth.

Representative numbers are useful. For \(\vinfearth=\SI{2}{\kms}\), one has \(\fg\simeq 32.4\), \(v_{\mathrm{imp}}\simeq \SI{11.38}{\kms}\), and \(\qspec\simeq \SI{64.7}{MJ\,kg^{-1}}\). For \(\vinfearth=\SI{20}{\kms}\), the focusing factor drops to \(\fg\simeq 1.31\), while \(v_{\mathrm{imp}}\simeq \SI{22.92}{\kms}\) and \(\qspec\simeq \SI{262.7}{MJ\,kg^{-1}}\). The point is not the size of the bulk energies by themselves, but the local pressure-temperature history of the buried microenvironment.

The lesson is twofold. Low-\(\vinfearth\) Earth encounters are geometrically favored, and low \(\vinfearth\) reduces the incremental dynamical load beyond Earth's unavoidable gravitational acceleration; yet even the most favorable Earth encounters still occur at meteorite-like speeds because Earth's gravity sets a floor under \(v_{\mathrm{imp}}\). What matters biologically is therefore not the total collision energy, but the local heating and pressure history of the buried microenvironment.

Atmospheric entry primarily ablates and heats an outer shell. Re-entry studies indicate that the outer several millimeters are strongly heated, while biologically relevant sterilization depths for small carriers are more plausibly of order \(2\) to \(5\) cm \cite{fajardo2005,delatorre2010}. Entry is thus a size filter, not a universal sterilizer. The transport chain remains open only for cargo buried at least centimeters deep, and preferably much deeper when long-flight radiation is relevant.

Final impact is less well constrained experimentally than launch shock or vacuum exposure, because survival depends strongly on target compliance: deep water, soft sediment, fragmentation during descent, and oblique low-mass impacts can all reduce the terminal stress seen by a buried interior \cite{delatorre2010,dartnell2011}. The conservative statement is therefore limited but important: Earth arrival is an additional low-probability filter, not an automatic sterilization event for deeply buried interiors.

\section{Transfer beyond the Solar System}
\label{sec:extrasolar}

\subsection{Birth-cluster transfer versus Galactic-field transfer}

Modern interstellar panspermia calculations distinguish sharply between the Sun's birth cluster and the later Galactic field. In the birth cluster, stars are closer together and relative speeds are smaller; capture is therefore less suppressed. Adams and Napier estimate an Earth-seeding expectation of roughly \(3\times10^{-5} f_{\mathrm{seed}}\) for the birth-cluster channel, compared with about \(5\times10^{-10} f_{\mathrm{seed}}\) for Galactic-field transfer and about \(3\times10^{-14} f_{\mathrm{seed}}\) for direct Earth impact from interstellar rocks \cite{adams2022}. Even the optimistic birth-cluster value is far below unity.

That analysis also reveals a second temporal bottleneck: if life could not arise on the donor before the Sun left its birth cluster, then the expected number of successful Earth-seeding events collapses even further \cite{adams2022}. Independent earlier modeling reached a closely related conclusion: outside the birth-cluster phase, potentially life-bearing extrasolar ejecta are not expected to have reached Earth before life already appeared here \cite{valtonen2009}. These results are difficult to evade because they follow from basic kinematics and timing rather than from an unusually pessimistic biological model.

The timing constraint can be stated more sharply. If life could not arise on the donor before cluster dispersal, the expected number of successful Earth-seeding events falls to about \(10^{-9} f_{\mathrm{seed}}\) \cite{adams2022}. More optimistic Galaxy-scale exchange models exist \cite{ginsburg2018,gobat2021}, but they do not overturn the Earth-specific conclusion: rocky bodies may traverse the Galaxy, yet Earth seeding remains strongly suppressed once one conditions on early arrival, Solar-System capture, and biological viability after long radiation exposure.

\subsection{Capture kinematics are decisive}

The capture problem is fundamentally a low-velocity problem in the heliocentric incoming speed \(\uinf\). Solar-System capture of interstellar objects is dominated by \(\uinf\lesssim \SI{4}{\kms}\) \cite{dehnen2022}. Galactic velocity dispersions are much larger than this capture-favored regime. Even before biological survival is considered, generic interstellar objects therefore rarely enter the kinematic window in which capture by the giant-planet architecture is efficient.

Capture kinematics are thus the decisive additional extrasolar filter. Once the low-\(\uinf\) requirement is compounded with the long-duration shielding requirement highlighted by Fig.~\ref{fig:dmin}, extrasolar hard panspermia to Earth becomes a conjunction of two already-suppressed events. More optimistic exchange models at Galactic scale remain physically interesting \cite{ginsburg2018,gobat2021}, but they do not change the Earth-specific conclusion because they do not remove the combined capture, timing, Earth-interception, and post-exposure viability penalties. 

\subsection{Why intergalactic hard panspermia is effectively negligible}

For natural intergalactic hard panspermia, even an optimistic lower-bound travel-time example is enough to show the scale of the problem. Consider a favorable halo-scale or very-near-neighbor case with \(d\sim100~\mathrm{kpc}\) and \(v\sim300~\kms\). The ballistic travel time is
\begin{equation}
t\sim \frac{d}{v}
\simeq 3.26\times10^{8}~\yr .
\label{eq:intergalactic_time}
\end{equation}
This is not a representative generic galaxy-to-galaxy distance; it is a deliberately favorable intergalactic-scale lower bound.

That timescale is already hostile to present-day terrestrial-like biochemistry. But the deeper problem is capture: intergalactic objects arrive far above the low-\(\uinf\) regime favored by Solar-System capture \cite{dehnen2022}. The chain therefore fails twice. First, the flight time pushes the biology into a radiation regime beyond the defensible end of Eq.~(\ref{eq:drad}). Second, the incoming speed lies orders of magnitude above the Solar-System capture-favored regime. Since even this favorable lower-bound example fails both the survival and capture filters, ordinary intergalactic transfer is still more strongly excluded.

A recent \textit{`Oumuamua}-motivated reanalysis reaches a closely related Earth-specific conclusion from updated interstellar number densities, placing an optimistic upper bound of order \(10^{-5}\) on the probability that biologically active interstellar ejecta seeded Earth before the earliest evidence for life, even under generous assumptions for shielding and establishment \cite{cao2024}. Thus the intergalactic result does not depend on a single pessimistic biological assumption; it follows from the joint flight-time and capture filters.

\section{Transit survival and carrier architecture}
\label{sec:biology}

\subsection{Surface biology is disfavored; endolithic packaging is favored}

The packaging problem is not cosmetic. Exposed cells are poor candidates because they receive direct solar UV during transit and experience the strongest heating during atmospheric entry. The most realistic passengers are endolithic or shallow-subsurface organisms that already inhabit pores, cracks, evaporites, or regolith grains before the ejection event. Lithopanspermia therefore selects for a particular biosphere architecture: if it occurs at all, it most naturally transports rock-inhabiting microorganisms rather than open-surface phototrophs or macroscopic organisms \cite{delatorre2010,dartnell2011,cottin2017}.

The biological benchmarks used below should be interpreted as survival envelopes for known resistant modern organisms and communities, not as a claim that early martian or Hadean terrestrial life had the same physiology as \textit{D.~radiodurans}, bacterial spores, or lichens. They define physically plausible upper-envelope constraints for dormant, shielded terrestrial-like biochemistry. If early life was less radiation-tolerant, less repair-capable after rehydration, or confined to shallow surface habitats, the hard-panspermia survival factors would be smaller than the fiducial estimates used here.

The relevant stresses are not interchangeable. Shock pressure and pulse duration dominate launch; peak temperature and heating duration dominate entry; accumulated dose and dose rate dominate long transfer; and dehydration, chemical alteration, and mechanical fragmentation can remove otherwise shielded cargo. The screening model responds by keeping pressure, burial depth, exposure time, encounter velocity, and material state as separate variables. Chemical alteration of the carrier--for example oxidation, radiolysis, salt or clay dehydration, and loss of volatile pore fluids--is not assigned a universal probability because it is lithology and history dependent. It is represented through \(\mumat\), the survival envelopes, and \(P_{\mathrm{est}}\). A carrier that satisfies the radiation depth requirement but fails the shock, chemistry, fragmentation, or terminal-impact gate is not counted as a viable hard-panspermia carrier.

\subsection{Vacuum, desiccation, and UV are survivable for years, not indefinitely}

Space-exposure data are now strong enough to support a very specific statement. Vacuum and desiccation are not automatically fatal to the most resistant microbes, provided UV is attenuated. Tanpopo showed that dried \textit{D.~radiodurans} pellets at least \(500\,\mu\mathrm{m}\) thick survived three years in space. Extrapolation of the measured survival curves implies that \(1\) mm pellets correspond to survival times of about \(2\) to \(8\) years in interplanetary space \cite{kawaguchi2020}. That is highly relevant for the rare shortest Mars-to-Earth transfers. It does not validate long interstellar transfer of naked cells.

\subsection{Ionizing radiation dominates beyond short flights}

For longer flights, ionizing radiation dominates the survival budget. Radiation-transport modeling summarized by Dartnell shows two nonintuitive but robust effects: rocks smaller than about \(60\) cm can have enhanced central dose because secondary cascades increase the dose in the interior, whereas boulders with diameter \(\gtrsim 2.6\) m begin to provide genuinely effective external shielding \cite{dartnell2011}. At burial depths of order \(1\) m, sterilizing doses accumulate on million-year timescales; at shallower depths and smaller sizes, the same total dose is reached much earlier \cite{mileikowsky2000,dartnell2011}.

This yields a practical hierarchy:
\begin{enumerate}
\item For \(t_{\mathrm{flight}} \lesssim 10\) yr, UV shielding and desiccation resistance dominate; millimeter-scale aggregates or centimeter-scale rock can suffice.
\item For \(t_{\mathrm{flight}} \sim 10^2\) to \(10^4\) yr, centimeter-to-decimeter lithic shielding becomes increasingly attractive.
\item For \(t_{\mathrm{flight}} \sim 10^6\) yr, meter-class shielding becomes the natural requirement.
\item For \(t_{\mathrm{flight}} \gtrsim 10^8\) yr, hard panspermia with contemporary Earth-like biochemistry is extremely difficult to defend quantitatively.
\end{enumerate}
That scale separation is the main reason Solar-System lithopanspermia remains plausible while interstellar and intergalactic hard panspermia do not.

\subsection{From flight time to required carrier architecture}

The transport filters can be combined into a practical engineering statement. Entry on Earth sets a nearly time-independent burial-depth floor \(\dent\), whereas radiation survival imposes a flight-time dependent shielding depth \(\drad(t_{\mathrm{fl}})\). The conservative lower envelope is therefore Eq.~(\ref{eq:dmin}). Figure~\ref{fig:dmin} compiles that envelope from the literature summarized in Table~\ref{tab:keyfilters}. It is not a universal constitutive law. Rather, it is a reviewer-auditable scenario-ranking tool that makes the scale hierarchy immediately visible.

The figure encapsulates a point that is easy to blur in narrative discussions. Fast Mars-to-Earth transfer is the only hard-panspermia channel that remains compatible with carrier sizes that are simultaneously plausible for natural ejection, compatible with entry survival, and not already excluded by long-duration radiation. Typical martian-meteorite transfer remains possible only for large, deeply buried carriers. Extrasolar channels inherit both the radiation burden of long flights and the Solar-System capture penalty associated with large \(\uinf\), which is why their expected Earth-seeding counts are so small.

\subsection{A survival-weighted viability measure for the low-shock spall population}
\label{sec:survivalweighted}

A persistent weakness of purely stage-based discussions is that they answer only whether \emph{some} sufficiently large carrier might survive. For hard panspermia, a better quantity is the fraction of the biologically loaded low-shock ejecta population that remains protected once the required burial depth is imposed. To make that idea explicit, consider a phenomenological low-shock carrier spectrum
\begin{equation}
\frac{dN}{dR}\propto R^{-\beta},
\qquad
R_{\min}\le R\le \Rmaxsh,
\label{eq:spectrum}
\end{equation}
where \(R\) is carrier radius, \(R_{\min}\) is the minimum carrier radius of interest, \(\Rmaxsh\) is the largest lightly shocked spall fragment available to the biologically viable ejecta tail, and \(\beta\) is a steep-spectrum index. Eq.~(\ref{eq:spectrum}) is not meant as a universal fragmentation law; its role is to isolate the biologically decisive competition between the required protected depth \(\dmin(t_{\mathrm{fl}})\) and the maximum lightly shocked carriers actually available.

The choice \(\beta=3.5\) used for the fiducial curves is a steep-spectrum
sensitivity case motivated by collisionally evolved fragment populations,
not a measured universal martian spall law \cite{obrien2003,housen2011}.
The conclusions below therefore depend on the relative scale of
\(d_{\min}(t_{\mathrm{fl}})\) and \(\Rmaxsh\), not on treating
\(\beta=3.5\) as unique.

As a fiducial volume-loaded endmember, assume that once a viable fragment exists the biological cargo is distributed through the fragment volume. This is appropriate for an idealized endolithic population occupying pore space and microfractures throughout the lightly shocked carrier, but it is not a conservative assumption for a shallow surface or near-surface biosphere. The surviving buried-volume fraction of a carrier of radius \(R\) after a flight time \(t_{\mathrm{fl}}\) is then
\begin{equation}
f_{\mathrm{bur}}(R,t_{\mathrm{fl}})=
\Theta\big(R-\dmin(t_{\mathrm{fl}})\big)
\Big[1-\frac{\dmin(t_{\mathrm{fl}})}{R}\Big]^3,
\label{eq:fbur}
\end{equation}
where \(\Theta\) is the Heaviside step function. Equation~(\ref{eq:fbur}) simply states that only the interior sphere remaining beneath a sacrificial shell of thickness \(\dmin\) is biologically protected. In this sense the fiducial model is best interpreted as an optimistic upper envelope for shallow-source biospheres and as a useful benchmark for genuinely endolithic or fracture-pervasive biological loading.

The survival-weighted buried-volume fraction of the low-shock ejecta population is
\begin{equation}
\Fbur(t_{\mathrm{fl}};\beta,R_{\min},\Rmaxsh)
=\frac{\displaystyle \int_{R_{\min}}^{\Rmaxsh}
R^{3-\beta}
 f_{\mathrm{bur}}(R,t_{\mathrm{fl}})\,dR}
{\displaystyle \int_{R_{\min}}^{\Rmaxsh}
R^{3-\beta}\,dR}.
\label{eq:Fbur}
\end{equation}
The weighting by \(R^3\) reflects the fact that, at fixed colonization density, biological cargo scales with fragment volume. Eq.~(\ref{eq:Fbur}) therefore asks the physically relevant population-level question: after the joint entry-and-radiation requirement is applied, what fraction of the biologically loaded low-shock ejecta volume remains viable? [A closed-form evaluation of Eq.~(\ref{eq:Fbur}) is given in Appendix~\ref{app:analytic}.]

\begin{figure}[t]
\centering
\includegraphics[width=0.60\columnwidth]{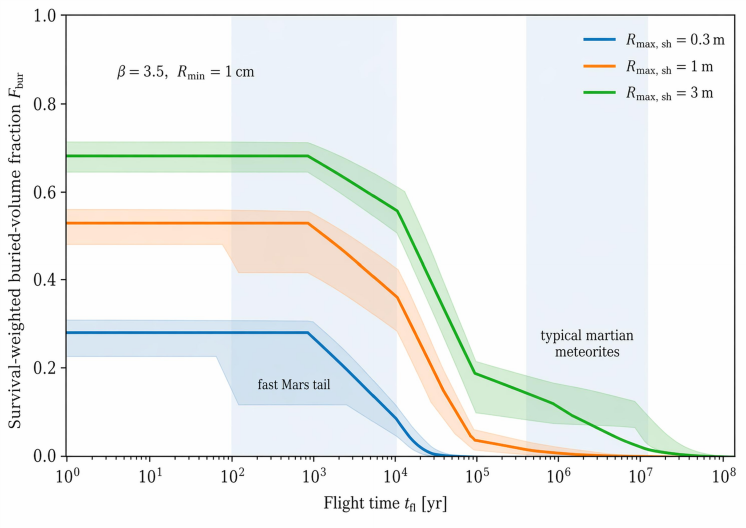}
\caption{Survival-weighted buried-volume fraction \(\Fbur\) from Eq.~(\ref{eq:Fbur}) for a representative steep ejecta spectrum with \(\beta=3.5\) and \(R_{\min}=\SI{1}{cm}\). The curves differ only in the maximum lightly shocked carrier radius \(\Rmaxsh\). Unlike Fig.~\ref{fig:dmin}, which is a carrier-level threshold, this panel is a population-level viability measure for the fiducial volume-loaded endmember; shell- or surface-biased biological loading is more strongly penalized. For \(\Rmaxsh\lesssim\SI{0.3}{m}\), the common Myr martian channel is effectively closed; for \(\Rmaxsh=\SI{1}{m}\), the fiducial model gives \(\Fbur\simeq1.36\times10^{-2}\) at \(d_{\min}=\SI{0.5}{m}\) and \(\Fbur=0\) as \(d_{\min}\rightarrow\SI{1}{m}\); only meter-to-multi-meter lightly shocked carriers keep long-flight viability appreciable.}
\label{fig:Fbur}
\end{figure}

It is useful to make the biological-loading assumption explicit. If the amount of biological cargo in a carrier of radius \(R\) scales as \(M_{\mathrm{bio}}(R)\propto R^{\gamma}\), then Eq.~(\ref{eq:Fbur}) generalizes to
\begin{equation}
\Fbur^{(\gamma)}(t_{\mathrm{fl}})
=
\frac{\displaystyle \int_{R_{\min}}^{\Rmaxsh}
R^{\gamma-\beta}
f_{\mathrm{bur}}(R,t_{\mathrm{fl}})\,dR}
{\displaystyle \int_{R_{\min}}^{\Rmaxsh}
R^{\gamma-\beta}\,dR}.
\label{eq:Fbur_gamma}
\end{equation}
The fiducial volume-loaded case used in Fig.~\ref{fig:Fbur} corresponds to \(\gamma=3\). A fracture- or surface-area-weighted distribution is closer to \(\gamma\simeq2\), and is more strongly penalized by the same protected-depth requirement because it places a larger fraction of the biological inventory in small or shallowly loaded carriers.

An even stricter shallow-shell model can be written directly. Let \(h_{\mathrm{bio}}\) be the thickness of an initially inhabited surface shell and define \([x]_+\equiv\max(x,0)\). Apart from the common factor \(4\pi/3\), the initially inhabited volume is
\begin{equation}
V_{\mathrm{bio}}^{\mathrm{sh}}(R;h_{\mathrm{bio}})
=
R^3-\bigl[R-h_{\mathrm{bio}}\bigr]_+^3,
\label{eq:Vbio_shell}
\end{equation}
whereas the part that remains deeper than the sacrificial shell \(d_{\min}\) is
\begin{equation}
V_{\mathrm{surv}}^{\mathrm{sh}}(R,t_{\mathrm{fl}};h_{\mathrm{bio}})
=
\Theta\bigl(R-\dmin\bigr)
\left[
\bigl(R-\dmin\bigr)^3
-
\bigl[R-h_{\mathrm{bio}}\bigr]_+^3
\right]_+ .
\label{eq:Vsurv_shell}
\end{equation}
The corresponding shell-loaded survival fraction is
\begin{equation}
f_{\mathrm{sh}}(R,t_{\mathrm{fl}};h_{\mathrm{bio}})
=
\frac{
V_{\mathrm{surv}}^{\mathrm{sh}}(R,t_{\mathrm{fl}};h_{\mathrm{bio}})
}{
V_{\mathrm{bio}}^{\mathrm{sh}}(R;h_{\mathrm{bio}})
}.
\label{eq:f_shell}
\end{equation}
Thus, if \(h_{\mathrm{bio}}\le d_{\min}\), a purely shallow-shell biosphere contributes no protected cargo in this model. This is the physically important contrast with the volume-loaded fiducial case: the martian channel is most favorable for deep endolithic or fracture-pervasive loading, and least favorable for purely surface-confined biology.

Figure~\ref{fig:Fbur} converts the protected-depth floor into a population-level viability penalty. For \(\Rmaxsh\lesssim \SI{0.3}{m}\), \(\Fbur\) is driven to zero once \(d_{\min}\) exceeds the largest lightly shocked carrier radius, as occurs in the common martian-meteorite range \(10^{5}\)--\(10^{7}\,\yr\). For the fiducial \(\beta=3.5\), \(R_{\min}=\SI{1}{cm}\), and \(\Rmaxsh=\SI{1}{m}\), the analytic expression gives \(\Fbur=0.48\) at \(d_{\min}=\SI{0.05}{m}\), \(0.31\) at \(d_{\min}=\SI{0.10}{m}\), \(1.36\times10^{-2}\) at \(d_{\min}=\SI{0.50}{m}\), and zero at \(d_{\min}=\SI{1}{m}\). Thus the common Myr martian channel is not mathematically closed for meter-scale lightly shocked carriers, but it is already suppressed by roughly two orders of magnitude once \(d_{\min}\) reaches half the maximum lightly shocked radius. Only if the lightly shocked spall population extends into the multi-meter regime does \(\Fbur\) remain appreciable at Myr flight times. The biologically privileged martian channel is therefore not Mars-to-Earth transfer in general, but the rare fast-transfer tail and/or unusually large lightly shocked carriers.

Three consequences follow directly. First, long-flight viability requires not merely that some large rock exist, but that the \emph{lightly shocked spall population} extend to radii comparable to the required protected depth. If \(\Rmaxsh\lesssim \SI{0.3}{m}\), then the common Myr martian-meteorite channel is biologically closed in the present model because \(\dmin\) eventually exceeds the largest lightly shocked carrier radius. If \(\Rmaxsh\sim \SI{1}{m}\), the long-flight channel is only marginally open; for \(\beta=3.5\), \(\Fbur\) at \(\dmin\sim \SI{0.5}{m}\) is only of order \(10^{-2}\), and it falls to zero as \(\dmin\) approaches \(\SI{1}{m}\). Only when \(\Rmaxsh\) reaches several meters does a substantial fraction of the low-shock biological volume remain viable over typical martian-meteorite exposure ages.

Second, survival weighting strongly boosts the biological importance of the rare fast tail. Rare \(10^{2}\)--\(10^{4}\) yr Mars-to-Earth transfers do not need to dominate the raw transfer count to matter biologically; they need only avoid the large suppression that Eq.~(\ref{eq:Fbur}) imposes on the much more common Myr channel.

Third, the conclusion strengthens for steeper spectra. Increasing \(\beta\) shifts more of the biological inventory into smaller fragments, which are precisely the carriers penalized most strongly by the burial-depth requirement. The curves in Fig.~\ref{fig:Fbur} are therefore conservative with respect to any surface-fragmentation regime dominated by small lightly shocked debris.

\section{Earth arrival: entry, heating, and terminal impact}
\label{sec:arrival}

\subsection{Why total collision energy is the wrong sterilization metric}

The concern about impact energy is physically justified but often misapplied. Even capture-favorable arrivals reach Earth with \(v_{\mathrm{imp}}\gtrsim \SI{11}{\kms}\), corresponding to bulk specific kinetic energies of order \(10^{2}\,\mathrm{MJ\,kg^{-1}}\) (Fig.~\ref{fig:arrivalkin}). But those are \emph{bulk} energies. They do not imply uniform heating or pressure throughout the carrier. The relevant survival variables are the thermal skin depth reached during entry and the pressure history transmitted to the buried cargo during final deceleration and impact.

A simple conductive estimate illustrates the point. If the outer surface is strongly heated for a duration \(t_{\mathrm{heat}}\), the corresponding thermal penetration depth is
\begin{equation}
\dth \sim 2 \sqrt{\alphath t_{\mathrm{heat}}},
\label{eq:dth}
\end{equation}
where \(\alphath\) is the thermal diffusivity of the rock. Using \(\alphath \sim 10^{-6}\,\mathrm{m^2\,s^{-1}}\) and \(t_{\mathrm{heat}} \sim 10^2\) to \(10^3\) s gives
\begin{equation}
\dth \sim 2 \text{--} 6~\mathrm{cm}.
\label{eq:dth_num}
\end{equation}
Equivalently, the diffusion time to a depth \(d\) is
\begin{equation}
\tau_{\mathrm{diff}}(d)\sim\frac{d^2}{\alpha_{\mathrm{th}}}
\simeq 2.5\times10^{3}\left(\frac{d}{0.05\,\mathrm{m}}\right)^2
\left(\frac{\alpha_{\mathrm{th}}}{10^{-6}\,\mathrm{m^2\,s^{-1}}}\right)^{-1}\mathrm{s}.
\label{eq:tdiff}
\end{equation}
Thus a \(10^2\)--\(10^3\) s entry-heating pulse can sterilize or devolatilize a centimeter-scale rind without equilibrating a decimeter-to-meter interior. This order-of-magnitude estimate is consistent with the experimental and meteoritic literature: outer shells are ablated or strongly heated, while sufficiently deep interiors can remain comparatively cool \cite{fajardo2005,delatorre2010,horneck2012,cottin2017}.

\subsection{Entry survival requires burial at least centimeters deep}

Entry is therefore not automatically sterilizing, but it is a serious size filter. Fajardo-Cavazos \etal\ showed that \textit{Bacillus subtilis} spores on artificial meteorites could survive a hypervelocity atmospheric-entry experiment in selected configurations \cite{fajardo2005}. By contrast, the STONE experiments demonstrated that shallowly exposed biological material on re-entry mounts did not survive the stagnation-point conditions tested \cite{delatorre2010}. The consistent inference is not that entry is fatal in all cases, but that viable cargo should be buried at least centimeters deep and preferably more deeply when the carrier is small or the entry is severe.

The survival-relevant timescales are stage-specific. Launch shock is a short pressure-temperature pulse, transit is a years-to-Myr radiation and desiccation exposure, atmospheric entry heats the outer shell over \(10^2\)--\(10^3\) s, and final deceleration can occur on \(10^{-4}\)--\(10^{-2}\) s scales depending on fragmentation and stopping length. These stresses cannot be collapsed into a single total-energy variable: thermal sterilization depends on time-temperature history, radiation survival depends on accumulated dose and shielding geometry, shock survival depends on peak pressure and pulse duration, and terminal impact survival depends on the stress actually transmitted to the buried interior.

\subsection{Impact deceleration, stopping length, and contact pressure}

The final deceleration environment is less well constrained than launch shock or vacuum exposure, but a useful engineering scaling can still be written down. Let \(\ell_{\mathrm{stop}}\) be the effective stopping length of the carrier during the terminal stage, after allowing for fragmentation and for the mechanical softness of the target medium. Then the characteristic deceleration and stopping time are
\begin{equation}
a_{\mathrm{stop}} \sim \frac{v_{\mathrm{imp}}^2}{2\ell_{\mathrm{stop}}},
\qquad
t_{\mathrm{stop}} \sim \frac{2\ell_{\mathrm{stop}}}{v_{\mathrm{imp}}}.
\label{eq:astop}
\end{equation}
For \(v_{\mathrm{imp}}\sim 12\,\kms\), one finds \(a_{\mathrm{stop}}\sim 7\times10^{7}(\ell_{\mathrm{stop}}/1\,\mathrm{m})^{-1}\,\mathrm{m\,s^{-2}}\). The terminal stage is therefore severe for compact impacts into hard targets, but it becomes less hostile when fragmentation, oblique entry, or impact into water or soft sediment increases the effective stopping length. The stress-communication time across a meter-scale silicate fragment is also short,
\begin{equation}
\tau_{\mathrm{ac}}\sim \frac{R}{c_s}
\simeq 2\times10^{-4}
\left(\frac{R}{1\,\mathrm{m}}\right)
\left(\frac{c_s}{5\,\mathrm{km\,s^{-1}}}\right)^{-1}\mathrm{s},
\label{eq:tac}
\end{equation}
so survival depends on the amplitude and duration of the transmitted pulse at the buried microenvironment, not merely on the carrier's pre-entry kinetic energy.

A contact-scale dynamic pressure estimate is still useful as an outer-envelope stress scale,
\begin{equation}
\pc \sim \rho_t v_{\mathrm{imp}}^2,
\label{eq:pc}
\end{equation}
where \(\rho_t\) is the density of the decelerating medium. For \(\rho_t \sim 10^{3}\) to \(3\times 10^{3}\,\mathrm{kg\,m^{-3}}\) and \(v_{\mathrm{imp}} \sim 11\) to \(20\,\kms\), the outer stagnation or contact zone reaches of order \(0.1\) to \(1\) TPa. That estimate is not the uniform stress seen by the carrier interior. Real interior stresses are reduced by ablation, fragmentation, sacrificial outer layers, and target compliance. The correct conclusion is not that impact survival is impossible, but that the terminal stage is another severe filter and that survival is most plausible for deeply buried cargo, especially when the effective stopping length is increased by favorable entry geometry or a mechanically soft target medium. Oblique entry, fragmentation that sheds a sacrificial outer shell, and impact into water, ice, or unconsolidated sediment all act in the same qualitative direction: they reduce the stress actually transmitted to the protected interior compared with the naive bulk-energy estimate.

\section{Integrated Earth-directed donor hierarchy}
\label{sec:donor}

The donor-ranking problem is not separate from the transport analysis above; it is the integrated use of that analysis. The relevant question is not whether a channel is conceivable in isolation, but whether its Earth-directed transport-survival weight remains non-negligible after the same launch, shielding, transfer, interception, entry, impact, and establishment filters are imposed on every donor class.

\paragraph*{Indigenous terrestrial origin.}

Earth pays no launch-from-donor penalty, no interplanetary or interstellar transfer penalty, no Solar-System capture penalty, and no atmospheric-entry penalty as an origin requirement. Its remaining unknowns are biological and geochemical rather than dynamical. That alone does not prove local origin, but it makes terrestrial abiogenesis the default null hypothesis against which all import channels must compete.

\paragraph*{Early Mars.}
If life was imported rather than locally originated, early Mars is the leading donor by a wide margin. The reason is cumulative rather than singular: Mars had early aqueous environments compatible with habitability; Mars-to-Earth lithic transfer is directly demonstrated; Mars has a low escape speed, which favors viable launch; and a rare short-transfer tail exists, which dramatically relaxes the radiation burden. No extrasolar donor presently matches that conjunction. The missing term in the Mars channel is biological rather than dynamical: no confirmed martian biosphere is known.

\paragraph*{Sibling birth-cluster systems.}
Sibling systems in the Sun's birth cluster remain the least implausible extrasolar donors because their relative velocities and separations were smaller than those of later Galactic-field systems. Even here, however, the expected Earth-seeding count remains far below unity, the donor must become biological before cluster dispersal, and long-duration survival still requires substantial shielding. This channel is physically interesting but not competitive with Mars for Earth's actual origin history.

\paragraph*{Galactic-field and intergalactic donors.}
Galactic-field donors fail because the low-\(\uinf\) Solar-System capture requirement and the radiation-limited survival requirement become unfavorable simultaneously. Intergalactic donors fail more decisively: incoming speeds are far above the capture-favored \(\uinf\) regime and flight times are so long that the required shielding becomes extreme before capture is even considered.

The resulting channel hierarchy is strongly ordered rather than binary. Transport physics does not prove indigenous terrestrial origin, but it places the alternatives into distinct viability classes. On current evidence, indigenous terrestrial origin remains the default inference; early Mars is the only external hard-panspermia channel whose Earth-directed kernel is not already negligible; sibling birth-cluster systems remain possible in principle but strongly suppressed; and Galactic-field or intergalactic hard panspermia is negligible for Earth's origin history.

\subsection{Relative channel weights versus absolute origin probabilities}
\label{sec:odds}

Absolute origin probabilities remain prior-dominated in the present problem, because the terrestrial
abiogenesis prior \(P_{\mathrm{abi},\oplus}\), the donor-life priors \(P_{\mathrm{life}}^{(i)}\), and the inhabited-ejecta
yields \(N_{\mathrm{ej,bio}}^{(i)}\) are not empirically known. What the present framework does determine is
the relative channel weight of each import mechanism once the Earth-directed transport-survival filter
is imposed.

We therefore define the unnormalized channel weights
\begin{equation}
W_\oplus \equiv P_{\mathrm{abi},\oplus}P_{\mathrm{est}}^{(\oplus)},
\qquad
W_i \equiv P_{\mathrm{life}}^{(i)}N_{\mathrm{ej,bio}}^{(i)}\mathcal{K}_{\oplus}^{(i)}P_{\mathrm{est}}^{(i\rightarrow\oplus)},
\label{eq:Wi}
\end{equation}
where \(W_\oplus\) is the weight for indigenous terrestrial origin and \(W_i\) is the corresponding
weight for donor class \(i\). The corresponding donor-to-Earth channel-weight ratio is
\begin{equation}
\mathcal{O}_{i/\oplus}\equiv \frac{W_i}{W_\oplus}
=
\frac{
P_{\mathrm{life}}^{(i)}N_{\mathrm{ej,bio}}^{(i)}\mathcal{K}_{\oplus}^{(i)}P_{\mathrm{est}}^{(i\rightarrow\oplus)}
}{
P_{\mathrm{abi},\oplus}P_{\mathrm{est}}^{(\oplus)}
}.
\label{eq:odds}
\end{equation}

Eq.~(\ref{eq:odds}) does not define a closed posterior. It defines a transport-ranked channel comparison: donor classes with \(\mathcal{K}_{\oplus}^{(i)}\ll 1\) are already noncompetitive before donor-biology
priors are assigned.  

The donor hierarchy is then not merely qualitative but thresholded. Galactic-field and intergalactic hard panspermia are already crushed at the transport level. Sibling birth-cluster channels remain strongly suppressed and require large source-side biological enhancement to compete. The only unresolved hard-panspermia comparison is therefore not Earth versus the Galaxy, but Earth versus early Mars, with the fast martian transfer tail singled out as the only external channel that avoids a large additional \(\Fbur\) penalty.

\subsection{Benchmark source-side compensation factors}
\label{sec:posterior_surrogate}

Eq.~(\ref{eq:odds}) is exact but not numerically closed, because the terrestrial abiogenesis prior \(P_{\mathrm{abi},\oplus}\), the donor-life prior \(P_{\mathrm{life}}^{(i)}\), the inhabited-ejecta yield \(N_{\mathrm{ej,bio}}^{(i)}\), and the establishment factor \(P_{\mathrm{est}}^{(i\rightarrow\oplus)}\) are not empirically known. The useful calculation is therefore not a posterior odds estimate, but a compensation-factor estimate: how large would the unresolved source-side terms have to be to offset a given transport penalty? Define
\begin{equation}
\mathcal{B}_i \equiv
\frac{
P_{\mathrm{life}}^{(i)}\,
N_{\mathrm{ej,bio}}^{(i)}\,
P_{\mathrm{est}}^{(i\rightarrow\oplus)}
}{
P_{\mathrm{abi},\oplus}\,
P_{\mathrm{est}}^{(\oplus)}
},
\label{eq:Bi}
\end{equation}
so that the donor-to-Earth channel-weight ratio factors as
\begin{equation}
\mathcal{O}_{i/\oplus}=\mathcal{B}_i\,\mathcal{T}_i .
\label{eq:OiTi}
\end{equation}
Here \(\mathcal{T}_i\) is the transport-controlled suppression factor. The symbol \(O_{i/\oplus}\) is therefore a ratio of unnormalized channel weights, not a closed Bayesian posterior odds ratio. If a channel weight were later interpreted as the mean number of independent successful seeding events, the corresponding event probability would be \(1-\exp(-W_i)\), reducing to \(W_i\) only in the rare-event limit \(W_i\ll 1\). The present paper compares the weights themselves and does not require that additional probabilistic closure.

For donor classes whose physics is already summarized by the Earth-directed kernel, we take
\begin{equation}
\mathcal{T}_i \equiv \mathcal{K}_{\oplus}^{(i)} .
\label{eq:Ti}
\end{equation}

For martian hard panspermia, the transport factor can be sharpened by explicitly separating the fast and slow transfer channels:
\begin{align}
\wp_{\mars}(t_{\mathrm{fl}})
&=
f_{\mathrm{fast}}\,
\wp_{\mathrm{fast}}(t_{\mathrm{fl}})
+
\bigl(1-f_{\mathrm{fast}}\bigr)\,
\wp_{\mathrm{slow}}(t_{\mathrm{fl}}),
\label{eq:wpmars_mix}
\\
\wp_{\mathrm{fast}}(t)
&=
\frac{
\Theta(t-t_{f_1})\Theta(t_{f_2}-t)
}{
t\,\ln(t_{f_2}/t_{f_1})
},
\qquad
(t_{f_1},t_{f_2})=(10^{2},10^{4})\,\yr,
\label{eq:wpmars_fast}
\\
\wp_{\mathrm{slow}}(t)
&=
\frac{
\Theta(t-t_{s_1})\Theta(t_{s_2}-t)
}{
t\,\ln(t_{s_2}/t_{s_1})
},
\qquad
(t_{s_1},t_{s_2})=(10^{5},10^{7})\,\yr.
\label{eq:wpmars_slow}
\end{align}

\begin{equation}
\overline{\Fbur}_{\mathrm{fast}}
\equiv
\int \wp_{\mathrm{fast}}(t_{\mathrm{fl}})
\,\Fbur(t_{\mathrm{fl}};\beta,R_{\min},\Rmaxsh)\,dt_{\mathrm{fl}},
\qquad
\overline{\Fbur}_{\mathrm{slow}}
\equiv
\int \wp_{\mathrm{slow}}(t_{\mathrm{fl}})
\,\Fbur(t_{\mathrm{fl}};\beta,R_{\min},\Rmaxsh)\,dt_{\mathrm{fl}}.
\label{eq:Fbur_fastslow}
\end{equation}

Thus the protected-volume part of the martian transport penalty is
\begin{equation}
\eta_{\mars}(f_{\mathrm{fast}},\Rmaxsh;\beta,R_{\min})
\equiv
f_{\mathrm{fast}}\,\overline{\Fbur}_{\mathrm{fast}}
+\bigl(1-f_{\mathrm{fast}}\bigr)\,\overline{\Fbur}_{\mathrm{slow}} .
\label{eq:etamars}
\end{equation}
To avoid double counting, we do not multiply this factor by the full kernel in Eq.~(\ref{eq:kearth}), because that kernel already contains \(S_{\mathrm{rad}}\) and \(S_{\mathrm{ent}}\). Instead define a residual martian transport kernel
\(\mathcal{K}_{\oplus,0}^{(\mars)}\) as the Mars-to-Earth kernel after the binary protected-volume requirement represented by \(\Fbur\) has been factored out. Equivalently, \(\mathcal{K}_{\oplus,0}^{(\mars)}\) collects the residual launch, escape, transfer, Earth-interception, terminal-impact, and any survival factors not already collapsed into \(\eta_{\mars}\). The effective martian transport factor is then
\begin{equation}
\mathcal{T}_{\mars,\mathrm{eff}}
=
\mathcal{K}_{\oplus,0}^{(\mars)}\,\eta_{\mars}.
\label{eq:Tmars_eta}
\end{equation}

The quantity \(\eta_{\mars}\leq 1\) is therefore not the full Mars-to-Earth transport kernel. It is the multiplicative loss associated only with the combined flight-time mixture and protected-volume requirement. Conditional on membership in the fast-transfer tail, \(\overline{\Fbur}_{\mathrm{fast}}\) can be order unity; the unconditional martian architecture factor remains \(\eta_{\mars}=f_{\mathrm{fast}}\overline{\Fbur}_{\mathrm{fast}}+(1-f_{\mathrm{fast}})\overline{\Fbur}_{\mathrm{slow}}\). Thus a rare fast tail is biologically privileged, but it is not automatically the whole Mars channel.

\begin{figure}[t]
\centering
\includegraphics[width=0.58\columnwidth]
{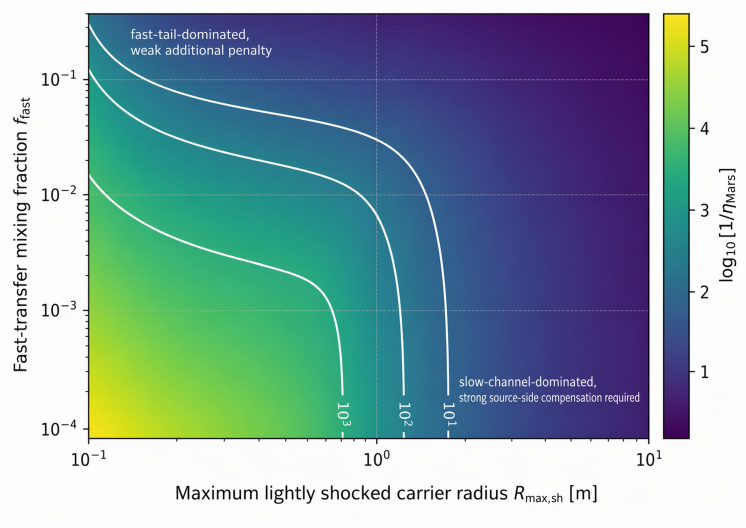}
\caption{Effective martian architecture-and-timing penalty implied by Eqs.~(\ref{eq:Fbur_fastslow})--(\ref{eq:etamars}). The color scale shows \(\log_{10}\eta_{\mars}^{-1}\), where \(\eta_{\mars}=f_{\mathrm{fast}}\overline{\Fbur}_{\mathrm{fast}}+(1-f_{\mathrm{fast}})\overline{\Fbur}_{\mathrm{slow}}\). The horizontal axis is the maximum lightly shocked carrier radius \(\Rmaxsh\), and the vertical axis is the fraction \(f_{\mathrm{fast}}\) of martian transfer events assigned to the biologically privileged fast tail. This is a sensitivity map for a factored transport penalty, not a posterior probability. Conditional on membership in the fast-transfer tail, the protected-volume penalty can be order unity; the unconditional martian architecture factor still includes the rarity of that tail through \(f_{\mathrm{fast}}\). Conversely, small \(\Rmaxsh\) and negligible \(f_{\mathrm{fast}}\) impose a large source-side compensation requirement before the residual Mars-to-Earth kernel \(\mathcal{K}_{\oplus,0}^{(\mars)}\) and the establishment factor are applied.
}
\label{fig:mars_transport_penalty}
\end{figure}

Figure~\ref{fig:mars_transport_penalty} makes the sensitivity explicit. If the inhabited low-shock ejecta population samples the fast-transfer tail appreciably, the protected-volume penalty can be small. If instead \(f_{\mathrm{fast}}\) is small and \(\Rmaxsh\) is sub-meter to meter-scale, \(\eta_{\mars}^{-1}\) can easily reach \(10^{2}\)--\(10^{3}\). Thus the relevant martian question is not simply whether Mars-to-Earth transfer occurs, but whether inhabited low-shock ejecta enter the rare short-transfer tail and/or occupy sufficiently large lightly shocked carriers.

For a factor \(\mathcal{T}_i\) that has been cleanly separated from source biology, the formal source-side compensation factor for parity with indigenous terrestrial origin is
\begin{equation}
\mathcal{B}_{i,\mathrm{crit}} \equiv \mathcal{T}_i^{-1},
\label{eq:Bcrit}
\end{equation}
and the corresponding threshold relative to the effective martian channel is
\begin{equation}
\mathcal{R}_{i/\mars}
\equiv
\frac{\mathcal{T}_{\mars,\mathrm{eff}}}{\mathcal{T}_i}.
\label{eq:Rimars}
\end{equation}

The Adams--Napier expectation values are not pure transport coefficients in the notation of Eq.~(\ref{eq:Ti}). Their seeding counts can be written schematically as
\begin{equation}
N_{\mathrm{seed},i}^{\mathrm{AN}}
=
N_{\mathrm{cap},i}\,
f_{B,i}\,
f_{\mathrm{surf},i}\,
f_{\mathrm{seed},i},
\label{eq:ANbookkeeping}
\end{equation}
where \(N_{\mathrm{cap},i}\) is the expected number of captured source rocks entering the relevant Solar-System/Earth-delivery channel, \(f_{B,i}\) is the biologically active source-rock fraction, \(f_{\mathrm{surf},i}\) is the fraction of delivered material that reaches a biologically relevant terrestrial environment, and \(f_{\mathrm{seed},i}\) is the post-arrival establishment probability. It is therefore too strong to identify the numerical coefficients quoted by \cite{adams2022} with \(\mathcal{T}_i\) itself.

We accordingly use those coefficients as \emph{benchmark composite ceilings},
\begin{equation}
\XAN_{\mathrm{cluster}}\sim 3\times10^{-5},
\qquad
\XAN_{\mathrm{field}}\sim 5\times10^{-10},
\label{eq:Textrasolar}
\end{equation}
defined by \(N_{\mathrm{seed},i}^{\mathrm{AN}}=\XAN_i f_{\mathrm{seed},i}\). The corresponding parity factors
\begin{equation}
\Bbench_{\mathrm{cluster},\mathrm{crit}}
\equiv
\XAN_{\mathrm{cluster}}^{-1}
\simeq
3.3\times10^4,
\qquad
\Bbench_{\mathrm{field},\mathrm{crit}}
\equiv
\XAN_{\mathrm{field}}^{-1}
\simeq
2.0\times10^9.
\label{eq:BcritNumbers}
\end{equation}
These \(\Bbench\) values are benchmark composite compensation factors, not measured source-biology requirements and not transport-only thresholds in the notation of Eq.~(\ref{eq:Ti}). They already mix capture, delivery, biologically active source-rock fraction, biologically relevant surface delivery, and the post-arrival establishment factor used by Adams and Napier. They should therefore be read as optimistic ceilings on the effectiveness of the corresponding extrasolar channels under those assumptions. Any \(f_{\mathrm{seed}}<1\), smaller biologically active source-rock fraction, smaller surface-delivery fraction, or additional radiation-survival loss increases the required compensation. The bookkeeping distinction does not change the transport conclusion: extrasolar hard panspermia remains strongly disfavored before donor-life priors are assigned, but the numbers \(3.3\times10^{4}\) and \(2.0\times10^{9}\) should be described as benchmark composite parity factors rather than precise biological thresholds.

Mars is different because its transport penalty is not represented by a single flight time. As Fig.~\ref{fig:mars_transport_penalty} shows, the protected-volume architecture penalty is controlled jointly by \(f_{\mathrm{fast}}\) and \(\Rmaxsh\). Conditional on membership in the fast-transfer tail, \(\overline{\Fbur}_{\mathrm{fast}}\sim O(1)\), so carrier architecture alone need not impose a large penalty. The unconditional martian channel, however, still carries the rarity of that tail through \(f_{\mathrm{fast}}\), as well as the residual kernel \(\mathcal{K}_{\oplus,0}^{(\mars)}\) and the establishment factor. If one instead asks about the common Myr martian-meteorite regime with \(\Rmaxsh\sim\SI{1}{m}\), then the protected-volume penalty alone is \(\eta_{\mars}\lesssim10^{-2}\), and the required compensation rises to \(\gtrsim10^{2}\). In that form the donor hierarchy is no longer merely verbal: Earth and the conditional fast Mars tail are the only hard-panspermia slices that avoid a large carrier-architecture penalty, while the full Mars channel remains sensitive to transfer timing, lightly shocked carrier size, biological loading depth, and establishment.

\begin{table}[t]
\caption{Transport-ranked parity and compensation factors for hard panspermia. For Mars, the listed values isolate the protected-volume architecture penalty \(\eta_{\mars}\) defined in Eq.~(\ref{eq:etamars}); the full martian transport factor is \(\mathcal{T}_{\mars,\mathrm{eff}}=\mathcal{K}_{\oplus,0}^{(\mars)}\eta_{\mars}\). The fast-tail row is conditional on membership in the biologically privileged short-transfer tail, whereas the unconditional Mars channel still carries the factor \(f_{\mathrm{fast}}\), the residual transport kernel, and the establishment factor. For extrasolar channels, \(\Bbench_{i,\mathrm{crit}}\) is a benchmark composite parity factor inferred from the Adams--Napier seeding coefficients; it is not a direct measurement of the transport-only factor \(\mathcal{T}_i\).}
\label{tab:Bcrit}
\setlength{\tabcolsep}{4pt}
\centering
\small
\begin{tabular}{lp{6.2cm}p{6.50cm}}
\toprule
Donor class & Transport statement & Parity factor / interpretation \\
\midrule
Conditional fast Mars tail
& \(\overline{\Fbur}_{\mathrm{fast}}\sim O(1)\); the unconditional contribution is \(f_{\mathrm{fast}}\overline{\Fbur}_{\mathrm{fast}}\)
& no large protected-volume penalty once the fast tail is selected; the full Mars channel still includes \(f_{\mathrm{fast}}\), \(\mathcal{K}_{\oplus,0}^{(\mars)}\), and \(P_{\mathrm{est}}^{(\mars\rightarrow\oplus)}\)
\\
Typical martian meteorite
& \(\Fbur\simeq 1.36\times10^{-2}\) at \(d_{\min}=\SI{0.5}{m}\) and \(\Fbur=0\) at \(d_{\min}=\SI{1}{m}\) for \(\Rmaxsh=\SI{1}{m}\)
& order-\(10^{2}\) or larger compensation is required for the protected-volume penalty alone
\\

Sibling birth-cluster system
& \(C_{\mathrm{cluster}}\sim3\times10^{-5}\) (benchmark composite ceiling)
& \(\widetilde{\mathcal{B}}_{\mathrm{cluster,crit}}\simeq3.3\times10^{4}\); least-suppressed extrasolar hard-panspermia channel, but still far below parity without source-side compensation
\\

Generic Galactic-field donor
& \(\XAN_{\mathrm{field}}\sim 5\times10^{-10}\) (benchmark composite ceiling)
& \(\Bbench_{\mathrm{field},\mathrm{crit}}\sim 2\times10^{9}\) \\
Intergalactic donor
& transport effectively closed
& no finite parity factor of practical relevance \\
\bottomrule
\end{tabular}
\end{table}

\section{Post-delivery establishment and soft panspermia}
\label{sec:establishment}

\subsection{Transport success is not biosphere success}

The Earth-side establishment factor \(\Pest\) is retained because
successful delivery is not equivalent to successful biosphere formation.
A viable arrival must still enter an early-Earth environment with usable
free energy, chemical retention, kinetic amplification, tolerable copying
error, sufficient coding capacity, and environmental persistence. We
therefore write
\begin{equation}
\Pest =
P\!\left[
\mathcal{G}_{\rm local}\cap \mathcal{G}_{\rm eco}
\,\middle|\,
\mathrm{delivery}
\right],
\label{eq:pest}
\end{equation}
where \(\mathcal{G}_{\rm local}\) denotes the local physicochemical
requirements for amplification and replication, and \(\mathcal{G}_{\rm eco}\) denotes the longer-term ecological and planetary persistence requirements. The present paper does not estimate \(\Pest\). It appears only to emphasize that transport success is a necessary but not sufficient condition for a successful hard-panspermia origin channel.  The broader gate-based biosphere framework in \cite{turyshev2026physicslife} provides a more general context.

\subsection{Soft panspermia is chemically more plausible}

The chemical case for extraterrestrial delivery is much stronger than the biological case. Carbonaceous meteorites contain a broad suite of prebiotically relevant compounds, including purine and pyrimidine nucleobases \cite{oba2022}. Recent analyses of Bennu samples further report abundant ammonia-rich soluble organics \cite{glavin2025} and bio-essential sugars \cite{furukawa2026}. Those results support repeated exogenous enrichment of early-Earth chemistry; they do not imply that life itself was imported.

Soft panspermia should therefore be treated as the delivery of organics, minerals, redox-active phases, and catalytic feedstock through the more forgiving chemical kernel \(\Kchem\), not as a weakened form of hard panspermia. Its relevance is that it can increase precursor inventories, catalytic surface area, mineral diversity, and redox disequilibrium without transporting a living cell. Nothing in the transport kernel alone constrains the later probability of evolving complex or intelligent life once a viable terrestrial lineage is established.

\begin{table*}[t]
\caption{Transport-ranked comparison of Earth-directed panspermia channels. For hard panspermia the listed assessments are conditioned on the existence of a donor biosphere and on later establishment
after arrival; the table therefore compares channels by the size of their Earth-directed transport penalty rather than by an absolute origin probability.}
\label{tab:scenarios}
\centering
\setlength{\tabcolsep}{3pt}
\renewcommand{\arraystretch}{1.05}
\begin{tabular}{@{}p{0.15\textwidth} p{0.12\textwidth} p{0.27\textwidth} p{0.23\textwidth} p{0.19\textwidth}@{}}
\toprule
Scenario & Characteristic flight time & Capture / encounter regime & Conservative carrier architecture for hard panspermia & Channel status \\
\midrule
Hard panspermia: fast Mars $\rightarrow$ Earth & \(10^{2}\) -- \(10^{4}\) yr & Solar-System transfer; no extrasolar capture penalty & Burial \(d\gtrsim 0.02\) -- \(0.05\) m; centimeter-to-decimeter lithic shielding is potentially sufficient & Physically plausible; biologically unproven \\
Hard panspermia: typical martian meteorite & \(3.5\times 10^{5}\) -- \(1.6\times 10^{7}\) yr & Solar-System transfer; no extrasolar capture penalty & Burial of order \(0.5\) -- \(1\) m and meter-class carriers become the natural requirement & Possible only for large, deeply shielded carriers; much less favorable than the fast Mars channel \\
Hard panspermia: sibling birth-cluster system & \(\gtrsim 10^{6}\) -- \(10^{8}\) yr & Requires low-\(\uinf\) Solar-System capture; expected seeding counts \(\sim 3\times10^{-5} f_{\mathrm{seed}}\) \cite{adams2022} & Multi-meter shielding plus early donor biosphere timing & Not excluded in principle, but strongly suppressed \\
Hard panspermia: generic Galactic-field donor & Long, typically \(\gtrsim 10^{7}\) yr & Capture rare because efficient Solar-System capture is dominated by \(\uinf \lesssim 4~\kms\) \cite{dehnen2022}; expected counts \(\sim 5\times10^{-10} f_{\mathrm{seed}}\) \cite{adams2022} & Multi-meter to larger shielding required, in a capture regime that is already disfavored & Negligible for Earth's origin \\
Hard panspermia: intergalactic donor & \(\sim 10^{8}\) -- \(10^{9}\) yr or longer & Incoming speeds of order hundreds of \(\kms\), far above the low-\(\uinf\) capture-favored regime; capture effectively fails & Required shielding becomes very large before capture is even considered & Effectively excluded \\
Soft panspermia: Solar-System organics & Not viability-limited & Solar-System delivery of organics and minerals is directly observed in meteorites and returned samples & No viable-cell shielding requirement; chemistry rather than viability dominates & Highly plausible as prebiotic enrichment \\
\bottomrule
\end{tabular}
\end{table*}

Table~\ref{tab:scenarios} summarizes the transport-ranked channel hierarchy: generic Galactic and intergalactic hard panspermia are already negligible at the transport level, sibling birth-cluster channels are strongly suppressed, and the only externally imported hard-panspermia channel that remains quantitatively serious is early Mars-to-Earth transfer.

\section{Discussion}
\label{sec:discussion}

The main technical result is a ranked hierarchy of transport penalties, not an absolute origin probability. Figure~\ref{fig:mars_transport_penalty} shows the internal martian sensitivity: the fast-transfer fraction and the maximum lightly shocked spall radius determine whether the protected-volume penalty is small or whether it adds two to three orders of magnitude of suppression. Conditional on a martian carrier entering the fast-transfer tail, the architecture penalty can be order unity; the unconditional Mars channel is instead \(\mathcal{K}_{\oplus,0}^{(\mars)}[f_{\mathrm{fast}}\overline{\Fbur}_{\mathrm{fast}}+(1-f_{\mathrm{fast}})\overline{\Fbur}_{\mathrm{slow}}]\), and therefore still carries the rarity of that tail. The extrasolar numbers in Table~\ref{tab:Bcrit} should be interpreted more conservatively: they are benchmark composite parity factors based on published seeding coefficients, not directly measured biological requirements. With that qualification, the hierarchy is still strongly ordered. Early Mars remains the only external hard-panspermia donor class that is not transport-negligible under current constraints; the common Myr martian-meteorite channel requires large lightly shocked carriers or additional source-side compensation; and birth-cluster and Galactic-field channels remain strongly suppressed by capture, timing, and radiation constraints.

The analysis constrains panspermia scenarios despite its simplifying assumptions because the filters are threshold-like. If \(d_{\min}(t_{\mathrm{fl}})>\Rmaxsh\), the low-shock protected-volume channel is closed in the fiducial model. If \(d_{\min}/\Rmaxsh\) is large but less than unity, Eq.~(\ref{eq:Fbur}) gives the corresponding reduction in biologically loaded carrier volume. If \(\uinf\) lies outside the low-speed capture window, extrasolar channels are suppressed before any biological prior is assigned. The framework therefore does not require knowing every organism-specific survival curve to identify which combinations of impactor material, transfer time, shielding depth, and encounter speed are plausible enough to justify detailed simulations.

\paragraph*{Sensitivity of the martian architecture penalty.}
The dominant internal sensitivity is the ratio \(d_{\min}/\Rmaxsh\), followed by the biological-loading exponent and the fast-transfer fraction. For the fiducial \(\beta=3.5\), \(R_{\min}=\SI{1}{cm}\), and \(\Rmaxsh=\SI{1}{m}\), increasing \(d_{\min}\) from \(0.10\) to \(\SI{0.50}{m}\) reduces \(\Fbur\) from \(0.31\) to \(1.36\times10^{-2}\), a factor of \(23\). At fixed \(d_{\min}=\SI{0.50}{m}\), increasing \(\Rmaxsh\) from \(1\) to \(\SI{3}{m}\) increases \(\Fbur\) from \(1.36\times10^{-2}\) to \(0.177\), a factor of \(13\). Changing the loading from volume-weighted to area-weighted is similarly important: using Eq.~(\ref{eq:Fbur_gamma}) with \(\gamma=2\) gives \(\Fbur\simeq1.6\times10^{-3}\) for \(d_{\min}=\SI{0.50}{m}\) and \(\Rmaxsh=\SI{1}{m}\). Thus the strong ordering of donor classes is robust, but the detailed Mars-to-Earth parity factor is controlled by three measurable or simulable quantities: the maximum lightly shocked spall radius, the depth distribution of biological loading, and the fraction of inhabited ejecta entering the fast-transfer tail.

Table~\ref{tab:sensitivity_budget} summarizes the broader uncertainty budget. The parameters most likely to change the detailed martian parity factor are \(\Rmaxsh\), \(\gamma\) or \(h_{\mathrm{bio}}\), \(d_{\min}(t_{\mathrm{fl}})\), and \(f_{\mathrm{fast}}\). The parameters most likely to change the extrasolar conclusion are qualitatively different: the low-\(\uinf\) Solar-System capture window, the early donor-biology timing requirement, and long-duration radiation survival. This separation is important because improving the martian channel requires changes in carrier architecture or transfer timing, whereas improving extrasolar hard panspermia requires simultaneous changes in capture, timing, and survival.

Three conclusions are robust. First, the Earth-directed transport-survival kernel is strongly scale-dependent. Within the inner Solar System, especially for Mars-to-Earth transfer, none of the major physical filters is obviously closed. Beyond the Solar System, several filters become simultaneously unfavorable. Second, the dominant bottleneck is flight-time dependent: short-transfer hard panspermia is limited mainly by launch shock, ultraviolet exposure, and entry burial depth, whereas Myr-scale transfer is dominated by cumulative ionizing dose and therefore by carrier architecture. Third, the biologically relevant variables are local and geometric: peak launch pressure, burial depth, accumulated dose, thermal penetration depth, encounter speed, and target compliance matter more than bulk kinetic energy.

The current ranking would change materially only if one of a small number of things turned out to be false. The first is biological: unambiguous evidence for a martian biosphere would immediately make the Mars channel much more consequential. The second is materials-dynamical: if the lightly shocked spall population extends routinely to meter-class radii, the long-flight martian channel becomes more open than the present conservative model allows. The third is radiobiological: if dormant viable cargo can survive \(10^{5}\)--\(10^{7}\,\mathrm{yr}\) without meter-class shielding, then the radiation filter would require major revision. The fourth is dynamical: if efficient Solar-System capture operates well outside the present low-\(\uinf\) regime, then extrasolar channels would need to be recalculated from the ground up. At present none of those revisions is supported strongly enough to overturn the current hierarchy.

The resulting inference is mixed rather than ideological. Exogenous delivery of organics and catalytic feedstock to the early Earth is highly plausible. For Earth's origin history, hard panspermia is transport-competitive only on Solar-System scales. Among nonterrestrial donors, early Mars remains the only one whose candidacy survives a stringent Earth-directed transport-survival analysis.

Transport physics constrains the identity and timing of a possible seed, not the later probability of evolving complex or intelligent life. Once a viable lineage is established, subsequent macroevolutionary outcomes are governed by ecology, contingency, and planetary feedbacks rather than by ejection, transfer, or entry mechanics \cite{turyshev2026physicslife}.

\paragraph*{Broader implication for life beyond Earth.}
The present result constrains the spread kernel, not the abiogenesis prior. If \(N_{\mathrm{hab}}\) denotes the number of habitable worlds in a Galactic population, \(f_{\mathrm{abi}}\) the probability of indigenous abiogenesis per habitable world, and \(K_{\mathrm{gal}}\) an effective galaxy-scale hard-panspermia transfer kernel, then the expected numbers of independently originated and secondarily seeded biospheres scale as
\begin{equation}
\langle N_{\mathrm{ind}}\rangle \sim N_{\mathrm{hab}} f_{\mathrm{abi}},
\qquad
\langle N_{\mathrm{seed}}\rangle \sim N_{\mathrm{src}} K_{\mathrm{gal}},
\label{eq:galactic_occurrence}
\end{equation}
with \(N_{\mathrm{src}}\) the number of already inhabited source systems. Earth establishes only that \(f_{\mathrm{abi}}\) is nonzero for at least one realized
planetary environment; it does not determine whether \(N_{\mathrm{hab}}f_{\mathrm{abi}}\)
is \(\ll 1\), \(O(1)\), or \(\gg 1\). The present analysis argues that \(K_{\mathrm{gal}}\) is very small for hard panspermia to Earth and, by the same transport logic, is not a plausible mechanism for rapid Galactic biological homogenization. That does \emph{not} imply that life is rare; it implies only that, if life is common, it must arise predominantly through multiple local origin events rather than through efficient intersystem contagion. Nothing in the present transport analysis constrains the later probabilities of complex multicellularity, intelligence, or galaxy-scale biological proliferation.

\section{Conclusions}
\label{sec:concl}

A semi-analytic, material-state-aware Earth-directed assessment of natural panspermia leads to five conclusions:

First, hard panspermia is sharply scale-dependent. For Earth's origin history, it is transport-competitive only on Solar-System scales, becomes strongly suppressed between stars, and is effectively negligible between galaxies.

Second, the governing state variables are local and material rather than bulk: peak launch pressure, burial depth, carrier lithology \(\mu\), density and porosity, shielding column \(\rho d\), cumulative radiation dose, heliocentric incoming speed \(\uinf\) for extrasolar capture, geocentric Earth-encounter speed \(\vinfearth\) for entry, thermal penetration depth during entry, fragmentation history, and target compliance at arrival.

Third, carrier architecture is set by the larger of the entry and radiation requirements. Earth entry imposes a burial floor of order \(2\)--\(5\) cm, whereas Myr-scale transit pushes the relevant hard-panspermia carriers toward the meter-class regime. Rare fast Mars-to-Earth trajectories are therefore biologically much more important than their raw dynamical rarity suggests.

Fourth, early Mars is the only external donor whose Earth-directed kernel is not already negligible. It combines early aqueous habitability, low escape speed, proven lithic exchange with Earth, and a nonzero fast-transfer tail. The missing ingredient is biological rather than dynamical: no confirmed martian biosphere is known.

Fifth, soft panspermia is chemically much more plausible than hard panspermia on large scales. Delivery of organics, catalytic minerals, and redox-active feedstock likely enriched the chemical context of early terrestrial abiogenesis even if life itself originated on Earth.

The paper therefore answers the terrestrial-origin question in ranked rather than binary form. Its purpose is not to solve the complete panspermia problem, but to identify the physical regimes in which transfer and survival are not already transport-negligible. Transport physics favors indigenous terrestrial origin as the null channel because Earth avoids launch from a separate donor, interplanetary or interstellar transfer, Solar-System capture penalties, atmospheric entry, and terminal-impact survival as origin requirements. This is not a posterior probability for terrestrial abiogenesis: the biological priors, inhabited-ejecta yields, and establishment probabilities remain unknown. If life was imported rather than locally originated, early Mars is the only hard-panspermia donor class that remains quantitatively serious after transport, timing, capture, shielding, entry, and establishment are imposed. Conditional fast Mars-to-Earth transfer can avoid a large carrier-architecture penalty, but the full martian channel still depends on the fast-tail fraction, the lightly shocked carrier-size distribution, biological loading depth, the residual transport kernel, and establishment. Extrasolar and intergalactic hard panspermia remain physically possible in the weak sense, but they are not competitive with either terrestrial abiogenesis or Mars-to-Earth transfer for Earth's actual origin history under present transport and survival constraints. Soft panspermia is a separate and much stronger claim: Earth was not chemically closed, and exogenous delivery of organics, minerals, and catalytic feedstock plausibly enriched the environment in which terrestrial life began.

\section*{Acknowledgments}
The work described here was carried out at the Jet Propulsion Laboratory, California Institute of Technology, Pasadena, California, under a contract with the National Aeronautics and Space Administration. 

\appendix

\section{Analytic form of the survival-weighted buried fraction}
\label{app:analytic}

Define
\begin{equation}
a\equiv \max(R_{\min},\dmin),
\qquad
b\equiv \Rmaxsh.
\end{equation}
If \(b\le \dmin\), then \(\Fbur=0\). Otherwise, Eq.~(\ref{eq:Fbur}) can be written as
\begin{equation}
\Fbur=
\frac{J_0-3\dmin J_1+3\dmin^2 J_2-\dmin^3 J_3}
{J_0^{\mathrm{tot}}},
\label{eq:Fanalytic}
\end{equation}
where
\begin{equation}
J_n=\int_a^b R^{3-\beta-n}\,dR,
\qquad
J_0^{\mathrm{tot}}=\int_{R_{\min}}^b R^{3-\beta}\,dR.
\end{equation}
For \(\beta\notin\{1,2,3,4\}\),
\begin{equation}
J_n=\frac{b^{4-\beta-n}-a^{4-\beta-n}}{4-\beta-n},
\qquad
J_0^{\mathrm{tot}}=\frac{b^{4-\beta}-R_{\min}^{4-\beta}}{4-\beta}.
\end{equation}
Logarithmic limits apply in the exceptional cases where the denominator exponent vanishes. Eq.~(\ref{eq:Fanalytic}) makes explicit why \(\Fbur\) is threshold-like in \(\Rmaxsh\): once the required protected depth approaches the largest lightly shocked carrier size, the surviving buried volume of the entire low-shock population collapses rapidly.

\section{Numerical validation of the screening estimates}
\label{app:numeric_audit}

This appendix provides numerical validation for the principal order-of-magnitude estimates used in the text, with a summary in Tables~\ref{tab:numerical_audit} and \ref{tab:Fbur_values}. The purpose is not to replace detailed radiation transport, impact hydrodynamics, or orbital integrations, but to make clear which numerical statements are direct consequences of the adopted scalings. 

All values in Tables~\ref{tab:numerical_audit} and \ref{tab:Fbur_values} are direct evaluations of the equations stated in the text. For example, \(\qspec=\frac{1}{2}v^2\) gives \(12.5\,\mathrm{MJ\,kg^{-1}}\) for \(v=\SI{5.0}{\kms}\) and \(62.7\,\mathrm{MJ\,kg^{-1}}\) for \(v=\SI{11.2}{\kms}\); Eq.~(\ref{eq:focusing}) gives \(\fg=32.4\), \(v_{\mathrm{imp}}=\SI{11.38}{\kms}\), and \(\qspec=64.7\,\mathrm{MJ\,kg^{-1}}\) for \(\vinfearth=\SI{2}{\kms}\); and Eq.~(\ref{eq:Fanalytic}) gives \(\Fbur=1.36\times10^{-2}\) for \(\beta=3.5\), \(R_{\min}=\SI{1}{cm}\), \(\Rmaxsh=\SI{1}{m}\), and \(d_{\min}=\SI{0.50}{m}\). The same analytic expression gives \(\Fbur=0.177\) for \(\Rmaxsh=\SI{3}{m}\) and \(d_{\min}=\SI{0.50}{m}\), and \(\Fbur=5.18\times10^{-2}\) for \(\Rmaxsh=\SI{3}{m}\) and \(d_{\min}=\SI{1.00}{m}\). These checks validate the order-of-magnitude hierarchy but do not replace detailed impact hydrodynamics, radiation transport, atmospheric-entry modeling, or orbital integrations.

\begin{table*}[t]
\caption{Numerical validation for the principal screening estimates. Values are rounded to the precision justified by the model assumptions.}
\label{tab:numerical_audit}
\centering
\setlength{\tabcolsep}{4pt}
\renewcommand{\arraystretch}{1.10}
\begin{tabular}{@{}p{0.24\textwidth} p{0.34\textwidth} p{0.32\textwidth}@{}}
\toprule
Quantity & Calculation & Result \\
\midrule
Mars escape specific energy &
\(\qspec=\frac12(5.0\,\mathrm{km\,s^{-1}})^2\) &
\(\qspec=\SI{12.5}{MJ\,kg^{-1}}\) \\
Earth escape specific energy &
\(\qspec=\frac12(11.2\,\mathrm{km\,s^{-1}})^2\) &
\(\qspec=\SI{62.7}{MJ\,kg^{-1}}\) \\
Low-\(\vinfearth\) Earth arrival &
\(\fg=1+11.2^2/2^2\), \(v_{\mathrm{imp}}=(2^2+11.2^2)^{1/2}\,\mathrm{km\,s^{-1}}\) &
\(\fg=32.4\), \(v_{\mathrm{imp}}=\SI{11.38}{km\,s^{-1}}\), \(\qspec=\SI{64.7}{MJ\,kg^{-1}}\) \\
High-\(\vinfearth\) Earth arrival &
\(\fg=1+11.2^2/20^2\), \(v_{\mathrm{imp}}=(20^2+11.2^2)^{1/2}\,\mathrm{km\,s^{-1}}\) &
\(\fg=1.314\), \(v_{\mathrm{imp}}=\SI{22.92}{km\,s^{-1}}\), \(\qspec=\SI{262.7}{MJ\,kg^{-1}}\) \\
Thermal penetration depth &
\(\dth=2(\alphath t_{\mathrm{heat}})^{1/2}\), \(\alphath=10^{-6}\,\mathrm{m^2\,s^{-1}}\), \(t_{\mathrm{heat}}=10^2\)--\(10^3\,\mathrm{s}\) &
\(\dth=0.020\)--\(0.063\,\mathrm{m}\), i.e. \(2\)--\(6.3\,\mathrm{cm}\) \\
Areal shielding column &
\(\Sigma=\rho d\), \(\rho=2.5\)--\(3.4\times10^3\,\mathrm{kg\,m^{-3}}\), \(d=0.05\) or \(0.50\,\mathrm{m}\) &
\(\Sigma\simeq1.3\)--\(1.7\times10^2\,\mathrm{kg\,m^{-2}}\) for \(5\) cm; \(1.3\)--\(1.7\times10^3\,\mathrm{kg\,m^{-2}}\) for \(0.5\) m \\
Terminal stopping estimate &
\(a_{\mathrm{stop}}=v_{\mathrm{imp}}^2/(2\ell_{\mathrm{stop}})\), \(v_{\mathrm{imp}}=\SI{12}{km\,s^{-1}}\), \(\ell_{\mathrm{stop}}=\SI{1}{m}\) &
\(a_{\mathrm{stop}}=7.2\times10^7\,\mathrm{m\,s^{-2}}\), \(t_{\mathrm{stop}}=1.7\times10^{-4}\,\mathrm{s}\) \\
Stress-communication time &
\(\tau_{\mathrm{ac}}=R/c_s\), \(R=0.1\)--\(1\,\mathrm{m}\), \(c_s=3\)--\(5\,\mathrm{km\,s^{-1}}\) &
\(\tau_{\mathrm{ac}}\simeq2\times10^{-5}\)--\(3\times10^{-4}\,\mathrm{s}\), a short transmitted-pulse timescale \\
Contact-pressure envelope &
\(\pc=\rho_t v_{\mathrm{imp}}^2\), \(\rho_t=10^3\)--\(3\times10^3\,\mathrm{kg\,m^{-3}}\), \(v_{\mathrm{imp}}=11\)--\(20\,\mathrm{km\,s^{-1}}\) &
\(\pc=0.12\)--\(1.2\,\mathrm{TPa}\), an outer-envelope contact-zone scale, not a uniform interior stress \\
Intergalactic flight time &
\(t=d/v\), \(d=\SI{100}{kpc}\), \(v=\SI{300}{km\,s^{-1}}\) &
\(t=3.26\times10^8\,\yr\) \\
Composite parity factors &
\((3\times10^{-5})^{-1}\), \((5\times10^{-10})^{-1}\) &
\(3.3\times10^4\), \(2.0\times10^9\) \\
\bottomrule
\end{tabular}
\end{table*}

For the fiducial buried-volume model, Eq.~(\ref{eq:Fanalytic}) gives the following representative values for \(\beta=3.5\) and \(R_{\min}=\SI{1}{cm}\):
\begin{table}[t]
\caption{Analytic values of \(\Fbur\) for representative protected depths and maximum lightly shocked carrier radii.}
\label{tab:Fbur_values}
\centering
\begin{tabular}{@{}cccc@{}}
\toprule
\(d_{\min}\) & \(\Rmaxsh=\SI{0.3}{m}\) & \(\Rmaxsh=\SI{1}{m}\) & \(\Rmaxsh=\SI{3}{m}\) \\
\midrule
\(\SI{0.05}{m}\) & \(0.204\) & \(0.480\) & \(0.676\) \\
\(\SI{0.10}{m}\) & \(5.97\times10^{-2}\) & \(0.309\) & \(0.546\) \\
\(\SI{0.50}{m}\) & \(0\) & \(1.36\times10^{-2}\) & \(0.177\) \\
\(\SI{1.00}{m}\) & \(0\) & \(0\) & \(5.18\times10^{-2}\) \\
\bottomrule
\end{tabular}
\end{table}
These values validate the qualitative hierarchy used in the main text while sharpening the wording: for \(\Rmaxsh=\SI{1}{m}\) and \(d_{\min}=\SI{0.5}{m}\), the fiducial volume-loaded model gives \(\Fbur=1.36\times10^{-2}\), not strictly less than \(10^{-2}\); the correct statement is that the surviving fraction is of order \(10^{-2}\) and falls to zero as \(d_{\min}\rightarrow\Rmaxsh\).


%

\end{document}